\documentclass[%
reprint,
 amsmath,amssymb,
 aps,
 pra,
longbibliography,
showkeys,
]{revtex4-1}

\usepackage[english]{babel}
\usepackage{graphicx}
\usepackage{epsfig}
\usepackage{color}
\usepackage{dcolumn}
\usepackage{bm}
\usepackage{hyperref}
\usepackage{siunitx}

\def\imag{{\rm i}}
\def\bra{\langle}
\def\ket{\rangle}
\def\Cm{C_{\rm m}}
\def\tsp{t^{\rm sp}}
\def\frate{f_0}
\def\Idc{I_{\rm DC}}

\def\stim{I}

\begin{document}

\title{Qualitative changes in spike-based neural coding and synchronization at the saddle-node loop bifurcation}

\author{Janina Hesse}
  \email{janina.hesse@bccn-berlin.de}
\author{Jan-Hendrik Schleimer}
  \email{jh.schleimer@hu-berlin.de}
  \thanks{shared first author}
\author{Susanne Schreiber}
 \homepage{www.neuron-science.de}
\affiliation{%
Institute for Theoretical Biology, 
Department of Biology,
Humboldt-Universit\"at zu Berlin,
\\
Philippstr. 13, Haus 4, 10115 Berlin\\
Bernstein Center for Computational Neuroscience Berlin
 }

\date{\today}


\begin{abstract} 

Information processing in the brain crucially depends on encoding
properties of single neurons, with particular relevance of the
spike-generation mechanism. The latter hinges upon the bifurcation
type at the transition point between resting state and limit cycle
spiking. Prominent qualitative changes in encoding have previously
been attributed to a specific switch of such a bifurcation at the
Bogdanov-Takens (BT) point. This study unveils another, highly
relevant and so far underestimated transition point: the saddle-node
loop bifurcation. As we show, this bifurcation turns out to induce
even more drastic changes in spike-based coding than the BT
transition. This result arises from a direct effect of the
saddle-node loop bifurcation on the limit cycle and hence spike
dynamics, in contrast to the BT bifurcation, whose immediate influence
is exerted upon the subthreshold dynamics and hence only indirectly
relates to spiking. We specifically demonstrate that the saddle-node
loop bifurcation $(i)$ ubiquitously occurs in planar neuron models with
a saddle-node on invariant cycle onset bifurcation, and $(ii)$ results
in a symmetry breaking of the system's phase-response curve. The
latter entails close to optimal coding and synchronization properties
in event-based information processing units, such as neurons. The
saddle-node loop bifurcation leads to a peak in synchronization range
and provides an attractive mechanism for the so far unresolved
facilitation of high frequencies in neuronal processing. The derived
bifurcation structure is of interest in any system for which a
relaxation limit is admissible, such as Josephson junctions and
chemical oscillators. On the experimental side, our theory applies to
optical stimulation of nerve cells, and reveals that these techniques
could manipulate a variety of information processing characteristics
in nerve cells beyond pure activation.
\end{abstract} 
\maketitle

\section{Introduction} 

Survival in a complex environment often demands nervous systems to
provide acute senses, intricate computations and speedy reactions
\cite{peron_spike_2009}, carried out in an economical fashion
\cite{laughlin_communication_2003}.  Efficient neuronal information
processing is hence considered an evolutionary favorable trait
\cite{niven_energy_2008}.  Information processing can be optimized on
the level of neuronal networks \cite{pouget_information_2000}, but also
on the level of single cells
\cite{koch_role_2000,schreiber_energy-efficient_2002,hesse_externalization_2015}.
For the latter, tuning the cell's voltage dynamics into a
regime that flexibly supports computational needs is
decisive.

In this study, we describe a point of drastic transition in neuronal
single-cell dynamics with consequences for information processing and
network synchronization. While the underlying bifurcation is not
unknown \cite{sato_changes_2014}, its consequences for neural
processing are severely underestimated. This article shows that the
transition in question turns out to be a ubiquitous feature in (planar)
type-I neuron models, known to describe various neurons, ranging from
isolated gastropod somata (Conner-Stevens model as stated in
Ref.~\cite{dayan_theoretical_2001}) to hippocampal neurons
\cite{traub_model_1991, wang_gamma_1996}.

Nerve cells are thought to encode information in the sequence of spikes
produced in response to their input. \emph{What} is encoded, crucially
depends on the specific mechanism of spike generation
\cite{ilin_fast_2013, hong_single_2007, brunel_firing-rate_2003,
wei_spike_2011, schleimer_coding_2009}. Different types of spike
generation were first classified by Hodgkin \cite{hodgkin_local_1948}
and later linked to particular bifurcations ruling the transition from
rest to spiking \cite{ermentrout_parabolic_1986,rinzel_analysis_1998}.
The detailed analysis of bifurcations in single neurons has so far
explained many of the complex responses that neurons show, \emph{e.g.},
bursting or rebound spiking
\cite{hindmarsh1984model,rinzel_analysis_1998,izhikevich_dynamical_2007}.
The mechanisms of other response properties, such as the sharp voltage
increase at spike onset, are still a highly debated topic
\cite{ilin_fast_2013,brette_sharpness_2013,naundorf_unique_2006}.

\begin{figure}
        \includegraphics[width=\columnwidth]{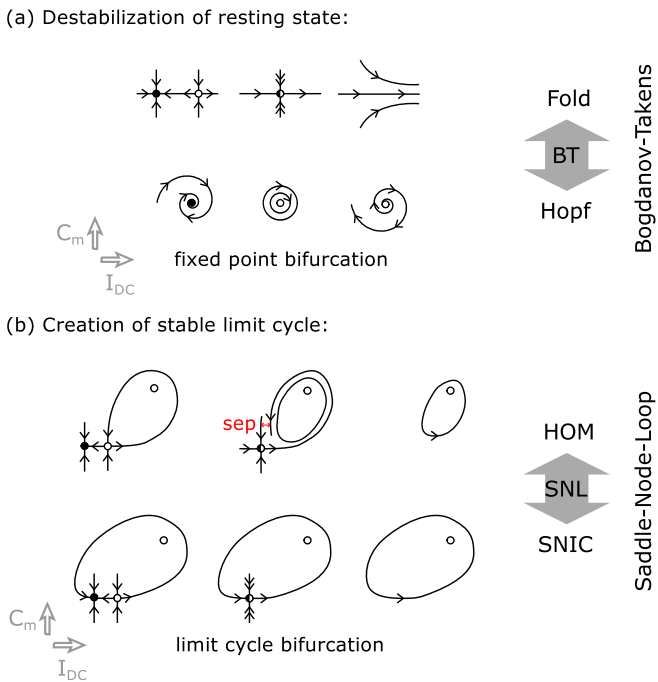}
        \caption{ \label{fig:bifOverview}
	{Color) The transition from rest
to spiking in response to an increase in input current $I_\text{DC}$ requires (a) that the resting state looses stability
(illustrated are fold and subcritical Hopf bifurcations) and 
(b) the creation of a limit cycle (illustrated are saddle homoclinic 
orbit (HOM) and SNIC bifurcations). The membrane capacitance $C_\text{m}$ 
allows to switch between these bifurcations. 
The separation function, $\text{sep}$,
marked in red, measures the distance between the stable 
and unstable manifold of the saddle. The overlap of both, \emph{i.e.},
$\text{sep} = 0$, results in a homoclinic orbit.
}}
    \end{figure}

\begin{figure} \includegraphics[width=\columnwidth]{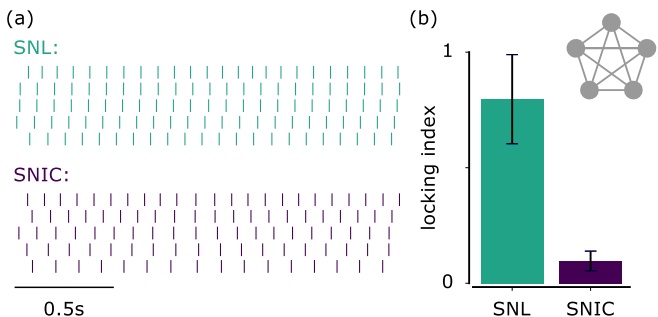}
      \caption{ \label{fig:syncDemo} {Color) (a) Spike raster plot of
          two small globally coupled network of 5 Wang-Buzsaki models,
          one close to the SNL bifurcation with $\Cm=1.47$ \si{\micro\farad}/cm$^2$,
          the other at a SNIC bifurcation with $\Cm=1$ \si{\micro\farad}/cm$^2$.
          Synaptic connections are modeled as $\delta$-perturbations of 0.35~mV.
          The frequency detuning of the neuron is approximately
          equally spaced between $5$ and $11$~\si\hertz. (b)
          The phase locking index was calculated between pairwise
          neurons $i$ and $j$ with phase $\phi_{i,j}$ as $\bra
          e^{\imag2\pi(\phi_i-\phi_j)}\ket$, where the brackets denote
          temporal averaging. Error bars denote standard deviations.}}
     \end{figure}

Recently, the ability of neurons to change the mechanism of spike
generation under physiological conditions has attracted the interest
of both theoreticians and experimentalists \cite{arhem_channel_2006,
stiefel_effects_2008, prescott_pyramidal_2008, phoka_new_2010}.
Attention was mostly directed at the transition between the two
traditional excitability types, which involve either a \emph{fold}
(saddle-node) or a \emph{Hopf} bifurcation
[Fig.~\ref{fig:bifOverview}(a)], along with their differential subthreshold
filtering properties
\cite{richardson_subthreshold_2003,schreiber_two_2009,
schreiber_subthreshold_2004,izhikevich_neural_2000}.  Here, we
investigate an alternative transition, which switches the spike onset
from a \emph{saddle-node on an invariant cycle} (SNIC) bifurcation to
a \emph{saddle homoclinic orbit} (HOM) bifurcation
[Fig.~\ref{fig:bifOverview}(b)]. This transition is organized by a
codimension-two bifurcation: the \emph{saddle-node loop} (SNL)
bifurcation \cite{schecter_saddle-node_1987} \footnote{The SNL
  bifurcation \cite{schecter_saddle-node_1987} is also known as
  \emph{saddle-node homoclinic orbit}, \emph{saddle-node noncentral
  homoclinic}, \emph{saddle-node separatrix-loop} bifurcation
\cite{izhikevich_dynamical_2007} or \emph{orbit flip} bifurcation
\cite{homburg_homoclinic_2010}.}. As we demonstrate, the SNL bifurcation causes an abrupt
change in phase-response curve, with far-reaching functional
consequences for spike-based coding. Moreover, the increase
in synchronization ability of individual cells observed at an SNL
bifurcation affects network synchronization [Fig.~\ref{fig:syncDemo}],
with potential relevance for various pathological conditions ranging
from epilepsy to Parkinson's disease \cite{jiruska_synchronization_2013,
schiff_towards_2010}. 

SNL bifurcations can occur with several bifurcation parameters, for
example the time constant of the gating kinetics
\cite{izhikevich_dynamical_2007}. In this study, we identify the
separation of time scales between voltage and gating dynamics as the
decisive bifurcation parameter, underlying the effect of other
parameters, such as capacitance or temperature. Starting at a SNIC
bifurcation in planar general neuron models, we demonstrate that a
variation in the separation of time scales provokes a generic sequence
of firing onset bifurcations. Compared to other bifurcation studies,
which rely on a \emph{local} unfolding of a codimension-three
bifurcation \cite{kirst_fundamental_2015,
pereira_bogdanovtakens_2015}, our approach proves the generic
bifurcation structure including appearance and ordering of
codimension-two bifurcations on a \emph{global} scale not restricted
to local analysis. The composed bifurcation diagram hence predicts the
behavior of a class of neurons over the whole range of time-scale
parameters, and thereby warrants a direct comparison with biological
neurons.

The organization of the paper reflects the intricate relation between
the dynamical and computational aspects of the SNL bifurcation.
Formally, this relation is establised via the phase-response
curve (PRC) \cite{ermentrout_relating_2007}. While
Sec.~\ref{sec.encoder} introduces the phase-response curve as
spike-based stimulus encoder, Sec.~\ref{sec.condBased} describes how it
can be identified from the limit cycle of a dynamical system.  With the
relation established, Sec.~\ref{sec.flip} proves that a symmetry
breaking of the phase-response
curve occurs at SNL bifurcations. The functional
consequences for coding and synchronization in spiking systems are
discussed in Sec.~\ref{sec.consequences}. The significance of these
consequences are perpetuated by the results in Sec.~\ref{sec.generic},
where we prove that SNL bifurcations generically occur in planar neuron
models.

\section{Spike-based information encoder}\label{sec.encoder}
The following two sections introduce phase oscillator models and
conductance-based neuron models, which we use to describe neuronal
dynamics on two different levels, the spike times, and the underlying
membrane voltage, respectively. In many nervous systems, the
integration of sensory stimuli from multiple modalities into an
appropriate behavioral response is achieved by translating the
impinging information into a series of spike times, $\{\tsp_k\}$, a
universal code for computation \cite{rieke_spikes:_1999}.  A single
neuron can be surmised as a \emph{spike time encoder} that maps input
signals, $\stim(t)$, to a spike train, $y(t)=\sum_k\delta(t-\tsp_k)$.
One way to formally identify a spike encoder from a biophysical model
of membrane voltage dynamics is the reduction to a phase equation
producing the same input-output mapping (\textit{i.e.}, an
\emph{input-output (I/O) equivalent} phase oscillator
\cite{lazar_population_2010}).  The I/O equivalent phase equation is
used throughout the paper to derive characteristics of neuronal
information processing, such as the maximal rate at which information
is transmitted, or the ability of neurons to synchronize their
activity.  Similar settings apply to many system ranging from
chronobiology \cite{bordyugov_tuning_2015} to pulsars
\cite{hilditch_introduction_2001}, where the only observations of the
complex dynamical system are pulse-like threshold crossings resulting
in an event time series.

The following analysis assumes tonic responses of a mean-driven neuron
\cite{schreiber_two_2009}, \textit{i.e.}, spikes are emitted with a
mean spike rate, $\frate$, in response to a constant mean stimulus intensity,
$\Idc$, and their occurrence is modulated by a time dependent,
zero-mean signal, $\stim(t)$, sufficiently weak to only shift spike
times. In this case, the mapping of input to spike times is given by
the \emph{phase-response curve} (PRC) of the neuron
\cite{lazar_functional_2014}. The PRC, $Z$, relates the timing of occurrence
of a weak perturbation to the resulting temporal advance or delay of
the following spike, $Z:\phi\mapsto\Delta\phi$.  The spike times
$\{\tsp_k\}$ correspond to the level crossings of the phase,
$\phi(\tsp_k)=k$ for $k\in\mathbb Z$, and their occurrence is governed
by the phase equation

\begin{equation}
    \label{eq.noisy_phaseoscil} 
    \dot\phi=\frate + Z(\phi)\stim(t) + \xi(t).
\end{equation}

The intrinsic noise $\xi(t)$ and the input $I(t)$ are assumed to be
zero-mean stochastic processes of different time scales: $\xi(t)$ is
white noise, $\langle\xi(0)\xi(\Delta t)\rangle=\sigma^2\delta(\Delta
t)$ and $I(t)$ is wide-sense stationary with correlations $\langle
I(0)I(\Delta t)\rangle=\varepsilon^2C(\Delta t)$.

To linear order, the I/O equivalent spike train is
$y(t)=\sum_k\delta(k-\phi(t))$. In the following, the mean spike rate $\frate$ in
response to the mean drive $I_{\text{DC}}$ and the PRC
$Z(\phi)$ are implicitly taken to be functions of the parameters of
the detailed neuron model introduced below.

\section{Conductance-based neuron model}\label{sec.condBased}

To investigate the mapping of input to spikes, our model neurons are stimulated by a constant
DC current and a zero-mean, time-varying stimulus, $I_\text{in} =
I_{\text{DC}}+\stim(t)$.  The dynamics of the membrane voltage $v$ follows a current balance
equation, $I_\text{in} = I_\text{cap} + I_\text{ion}$. The input
causes a capacitive current, $I_\text{cap} = \frac{\text{d} C_\text{m}
v}{\text{d} t}$ (with membrane capacitance $\Cm$) and an ionic
current,  $I_\text{ion} = I_\text{ion}(v, m_{i}, ...)$, which
is a function of $v$ itself and the open probability of ion channels
given by their \emph{gating variables} $m_{i}$.

Combined, this so-called \emph{conductance-based neuron model} entails a
dynamical system, $\dot{X} = F(X)+G(X,t)$, with the following
structure: 

\begin{equation} 
  \left(\begin{array}{c}
         \dot{v} \\ \dot{m}\\  ... \\
        \end{array}\right)
 = \left(\begin{array}{c}
   \frac{1}{C_\text{m}}
   (I_{\text{DC}} - I_\text{ion}(v, m, ...) + \stim(t))\\
\frac{m_{\infty}(v)-m}{\tau_{m}(v)} \\ ... \\
        \end{array}\right),
  \label{eq:dynSys} 
\end{equation} 

where the dot $\dot{}$ denotes the derivative by time, $F$ determines
the dynamics of the unperturbed system, and $G=\Cm^{-1}\stim(t)
{e}_{v}$ is some time-dependent voltage perturbation
(${e}_{v}$ represents the basis vector in voltage direction). The
dynamical variables consist of the voltage and the gating variables
such as $m$. The gating is typically modeled by first-order kinetics
(for details see
Appendix~\ref{sec.modelDef}). 

The input $I_{\text{DC}}$ acts as bifurcation parameter for the
bifurcations of both fixed point destabilization and limit cycle
creation [Fig.~\ref{fig:bifOverview}]. For our analysis, we focus on neuron models in which the
fixed point looses stability at a fold bifurcation.  To identify the I/O equivalent phase model in
Eq.~\eqref{eq.noisy_phaseoscil}, the PRC needs to be calculated for the
conductance-based model in Eq.~\eqref{eq:dynSys}.  From a dynamical
systems perspective, the PRC $Z$ is the periodic solution to the
adjoint of the first variational equation of the unperturbed
dynamics in Eq.~\eqref{eq:dynSys}, $\dot X = F(X)$,

\begin{equation}
    \label{eq.adjoint} 
    \frac{\text{d}Z}{\text{d}\phi}(\phi) =-J^\top(\phi)Z(\phi), 
\end{equation} 

where $\top$ denotes the matrix-transpose and
$J=\frac{{\partial}F}{{\partial}X}$ is the \emph{Jacobian} evaluated on the
limit cycle. To comply with Eq.~\eqref{eq.noisy_phaseoscil}, the PRC associated
with input current perturbations needs to be normalized as $Z(\phi)\cdot
F(\phi)=\frate/\Cm,\forall\phi$. The resulting relation between PRC and
parameters of the conductance-based neuron model allows us to consider coding
properties at different firing onset bifurcations. In the following, we use the
dynamics on the homoclinic orbit to infer PRC properties of the limit cycle
that arises from the homoclinic orbit, and, for convenience, we refer to the
limit cycle PRC as the PRC \emph{at} the limit cycle bifurcation (SNIC or SNL),
\textit{i.e.}, $Z_\text{SNIC}$ or $Z_\text{SNL}$.

\section{A flip in the dynamics alters the PRC symmetry at an SNL bifurcation} 
\label{sec.flip}

In a first step, we infer coding properties from the dynamics at firing onset
bifurcations, in particular around the SNL bifurcation. 
As bifurcations imply in general
qualitatively different dynamics \cite{kuznetsov_elements_2013}, limit
cycle
dynamics are expected to change at the switch in firing onset dynamics at the
SNL bifurcation. But what are the specific consequences for the way
neurons encode stimuli and synchronize in networks? To answer this
question, we start by discussing changes in limit cycle dynamics at the SNL
bifurcation. We then show that this also alters the PRC in such a
profound way that it has, in turn, drastic implications for the
resulting coding properties discussed in Sec.~\ref{sec.consequences}.
To this end, we use the tight relation between spike coding and PRC
[Eq.~\eqref{eq.noisy_phaseoscil}], as well as PRC and dynamics
[Eq.~\eqref{eq.adjoint}]. 
    
    \begin{figure} \includegraphics[width=\columnwidth]{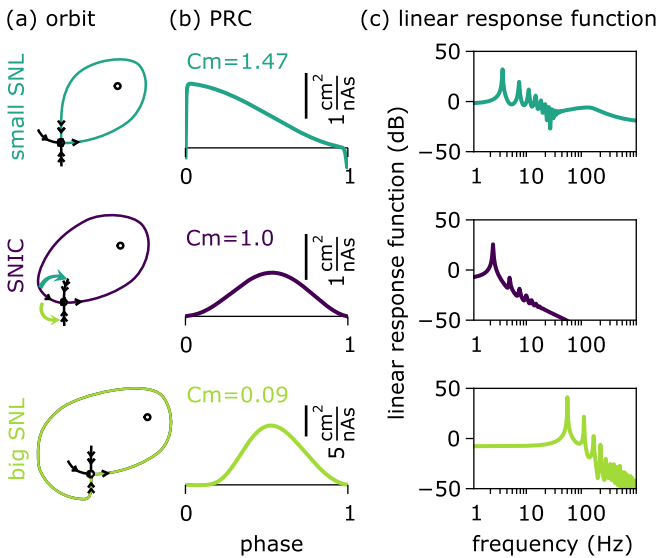}
      \caption{ \label{fig:figure1} {Color) Top to bottom: (a)
          Schematic illustration of the orbits at small SNL
          bifurcation, non-degenerated SNIC bifurcation, and big SNL
          bifurcation, with semi-stable (small single arrow) and
          strongly-stable manifold (double arrows).  These
          bifurcations occur in the Wang-Buzsaki model for
          $I_\text{DC} \approx 0.16$~\si{\micro\ampere}/cm$^2$,
          $C_\text{m} \approx [1.47,\, 1,\,
          0.09]$~\si{\micro\farad}/cm$^2$. 
          (b) and (c) show the
          associated phase-response curves and transfer 
          functions, for $I_\text{DC}$ 2\% above the fold
    bifurcation.  } } \end{figure}

\subsection{Orbit flip}
\label{sec.dynSNL}

We consider models with classical type-I excitability where the
transition from rest to repetitive firing
is marked by $(i)$ the elimination of the resting state in a
fold bifurcation, and $(ii)$ the existence of a limit cycle to which
the dynamics relax instead.  This limit cycle is born at a \emph{limit
cycle bifurcation}, which is in type-I neurons typically a SNIC
bifurcation. At a codimension-two SNL bifurcation, the limit cycle
bifurcation switches between a SNIC and a HOM bifurcation
[Fig.~\ref{fig:bifOverview}(b)]. The following, model-independent
analysis focuses on the \emph{small} SNL bifurcation that transitions
from a SNIC orbit to a small HOM orbit [Fig.~\ref{fig:figure1}(a)], because
it entails more drastic changes in PRC shape, as discussed later.  The
\emph{big} SNL bifurcation (transitioning to a big HOM orbit) will be
studied with numerical continuation [Sec.~\ref{sec.consequences}].

The limit cycle created at a HOM, SNIC or SNL bifurcation arises from a
homoclinic orbit to a saddle (HOM) or saddle-node (SNIC, and also SNL).
Under the assumption of sufficiently large limit cycle periods, the
slow velocity in the vicinity of these fixed points contracts the
dynamics such that limit cycle properties, \textit{e.g.}, period
or PRC, can be extracted from a linear approximation around the fixed
point.  

The linearized dynamics around the saddle-node
fixed point is given by its Jacobian. Assuming non-degeneracy, 
the Jacobian has a single zero eigenvalue,
associated with the \emph{semi-stable} manifold, while the other
eigenvalues are strictly negative (\emph{strongly-stable} manifolds).
Trajectories always leave the saddle-node along the semi-stable
manifold. When a trajectory loops around in a homoclinic orbit, it can
either re-approach the saddle-node along the same manifold (SNIC
bifurcation), or along the much faster, strongly-stable manifold (SNL
bifurcation). The approach of the saddle-node at an SNL
bifurcation flips from the semi-stable manifold to one of the strongly-stable
manifolds (hence \emph{orbit flip} bifurcation
\cite{homburg_homoclinic_2010}) [Fig.~\ref{fig:figure1}(a)].  For
neuron models, this flip can be induced by a scaling of the relative
speed in the voltage and gating kinetics [Fig.~\ref{fig:phase}].  When the saddle-node
disappears after the fold bifurcation, its remaining \emph{ghost}
still dominates the resulting limit cycle dynamics. The limit cycle
period drastically decreases around the SNL bifurcation
[Fig.~\ref{fig:figure2}(a), see also \cite{sato_changes_2014}], mainly
because of the separation of time scales between strongly-stable and
semi-stable manifold, which renders the approach along the
strongly-stable manifold much faster than the approach along the
semi-stable manifold. 

\subsection{PRC symmetry and Fourier modes} \label{sec.prcsym}

Numerical continuation of several neuron models shows that the PRC is
drastically altered at the SNL bifurcation.  Exemplified in
Fig.~\ref{fig:figure1}(b) for the Wang-Buzsaki model
[Sec.~\ref{sec.modelDef}], the symmetric PRC at a (non-degenerated)
SNIC bifurcation becomes increasingly asymmetric when an increase in
membrane capacitance tunes the model towards the SNL bifurcation. 
The strong asymmetry at the SNL
bifurcation directly affects the synchronization ability of the neuron
[see Sec.~\ref{sec.consequences}]. 

The sudden occurrence of PRC asymmetry at an SNL bifurcation can be
directly inferred from the orbit flip in the dynamics described in the
last section [Sec.~\ref{sec.dynSNL}]. The PRC peaks when the phase
reaches the ghost of the saddle-node, where the slow dynamics allow
infinitesimal perturbations to maximally advance phase. In the case of
the SNIC bifurcation, the same velocity governs approach and exit of
the ghost, both aligned with the semi-stable manifold [Fig.~\ref{fig:phase}, for details see
Appendix~\ref{sec.prcsymAppendix}].  The orbit flip to the strongly
stable manifold at the SNL bifurcation either decreases or increases
the time spent on the approach compared to exit for the small or big
SNL, respectively.  This in turn breaks the symmetry of the PRC at the
SNIC bifurcation by advancing or delaying the phase at which the
maximum of the PRC resides.

Neglecting the fast approach at the small SNL bifurcation, it seems as
if the flow of the limit cycle trajectory is directly injected at the
ghost. Because the exit dynamics at SNL and SNIC bifurcations are
similar, the PRC at the small SNL bifurcation appears as a rescaled
version of the second half of the PRC at the SNIC bifurcation,
$Z_\text{small SNL}(\phi)
\def\p{\setbox0=\vbox{\hbox{$\propto$}}\ht0=0.6ex \box0 }
\def\s{\vbox{\hbox{$\sim$}}}\mathrel{\raisebox{0.7ex}{
\mbox{$\underset{\s}{\p}$}}}Z_\text{SNIC}(0.5\phi+0.5)$. This
reasoning is supported by numerical continuation
[Fig.~\ref{fig:figure1}(b), Fig.~\ref{fig:phase}], 
and explains the observation that the limit
cycle period is approximately halved at the SNL
bifurcation [Fig.~\ref{fig:figure2}(a)]. 

The necessity of the PRC symmetry breaking at the SNL bifurcation can
also be seen from normal form theory.  For the SNIC bifurcation (and
the supercritical Hopf bifurcation), the PRC is a simple trigonometric
function, $Z_\text{SNIC}(\phi) \propto 1-\cos(2\pi\phi)$
($Z_\text{Hopf}(\phi) \propto \sin(2\pi\phi)$) \cite{brown_phase_2004}.
Approached from the SNIC, the small SNL bifurcation, however, registers
a sudden emergence of higher Fourier modes in the PRC.  On the
other side of the small SNL bifurcation, the canonical PRC at a small
HOM bifurcation is an exponential with some decay constant $\tau$,
$Z_\text{HOM}(\phi) \propto \exp(-\phi/\tau)$
\cite{brown_phase_2004,shaw_phase_2012}.  Hence, in contrast to the
trigonometric PRCs with a single Fourier mode at the SNIC or
supercritical Hopf bifurcations, the PRCs at HOM and small SNL
bifurcations have an infinite amount of Fourier modes.  This results in
Gibb's phenomenon if finite approximations are used.

The significant increase in PRC Fourier modes, as well as the
breaking in PRC symmetry, are generic properties of SNL bifurcations.
The consequences for coding are detailed in the next section
[Sec.~\ref{sec.consequences}].

    \begin{figure}
      \includegraphics[width=\columnwidth]{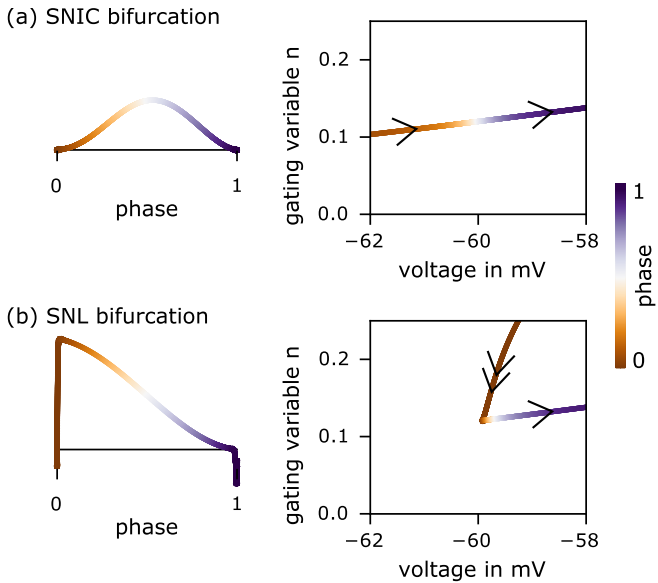} \caption{
        \label{fig:phase} {Color) Phase-response curve (left) and
          phase plot around the saddle-node (right) for (a) a
          non-degenerated SNIC bifurcation with
          $C_{\text{m}}=1$~\si{\micro\farad}/cm$^2$ and (b) a small
          SNL bifurcation with $C_\text{m} \approx
    1.47$~\si{\micro\farad}/cm$^2$ in the Wang-Buzsaki model with the
    limit cycle period fixed in both cases to 2~\si\hertz.  } }
  \end{figure}

\section{Spike-based coding and synchronization around SNL
bifurcations} \label{sec.consequences}

While the previous section established changes in PRC properties at
SNL bifurcations using dynamical system's theory, this section takes
up the computational perspective again, with the PRC as a spike-time
encoder.
From the PRC, the following paragraphs derive various measures to
probe the performance of an information encoding system.  The
pertinence of these measures for spike-based coding is discussed at
the end of this section.  We start by interpreting the spike-time
encoder as an information filter, which allows for the derivation of the
stimulus characteristics to which the neurons are particularly
sensitive [Sec.~\ref{sec.lrf}]. The transmission of information
through such a filter is quantified in Sec.~\ref{sec.info}. The
remaining paragraphs consider the ability of the neuron to align its
activity to an entraining stimulus, either in the context of
stochastic synchronization due to a common input
[Sec.~\ref{sec.reliability}], or by classical synchronization, for
example with the population activity [Sec.~\ref{sec.sync}]. All four
measures peak around the SNL bifurcation. Two factors are decisive:
The reduced limit cycle period at the SNL bifurcation and the PRC
symmetry breaking with the emergence of high frequency Fourier modes. 
Both occur generically at SNL bifurcations
[Sec.~\ref{sec.flip}], such that the consequences derived
in this section generalize to other information-processing systems beyond
neuroscience.
In particular, neurons close to an
SNL bifurcation behave radically differently from what is expected for
SNIC neurons that show traditional type-I excitability. 

    \begin{figure}
    \includegraphics[width=\columnwidth]{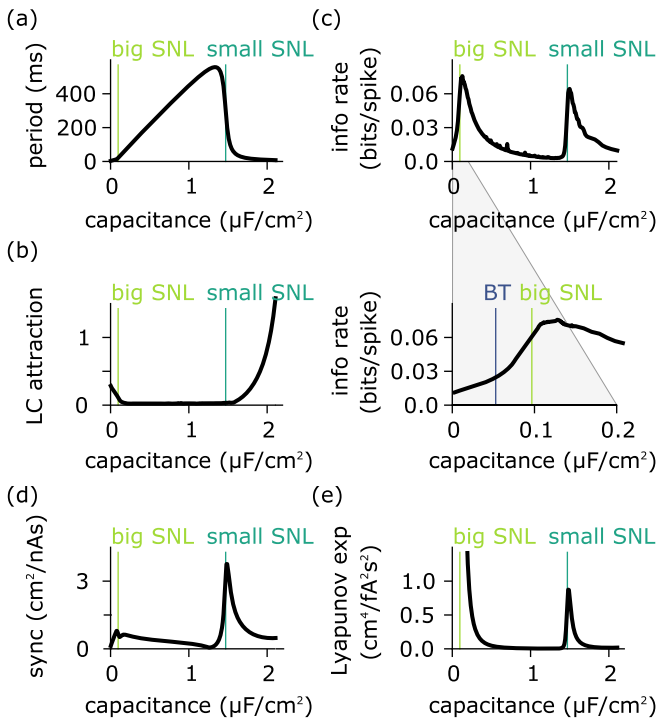} 
    \caption{ 
    \label{fig:figure2}
	{Color) Coding properties against the capacitance $C_\text{m}$
for the Wang-Buzsaki model with input fixed at 2\% above limit cycle
onset at $I_\text{DC} \approx 0.16$ \si{\micro\ampere}/cm$^2$. (a) Limit cycle
period. (b) Relative limit cycle (LC) attraction given by the ratio
of the limit cycle attraction time (inverse of the Floque exponent) and the period.
A small LC attraction supports the validity of the phase description
\cite{kuramoto_chemical_1984}. (c) Information rate as defined in
\cite{schleimer_coding_2009}, see Sec.~\ref{sec.info}, with zoom below. Note the maxima in
proximity of both SNL bifurcations. (d) Maximal amplitude of the odd
part of the PRC (entrainment range normalized by the coupling strength
assuming $\delta$-coupling; abbreviated sync). (e) Magnitude of the
Lyapunov exponent, $|\lambda|$, of the locking state to a time-varying
white noise stimulus.  } } 
    \end{figure}

\subsection{High-frequency transmission peaks around SNL points}\label{sec.lrf}

The natural environment is brimming with dispensable information from
which sensory systems filter out behaviorally relevant information
\cite{machens_representation_2001, fox_encoding_2010,
neiman_sensory_2011}. From a computational perspective, 
frequency band selectivity is given by the
first-order relationship between input and output, the \emph{linear
response function} (first-order Volterra expansion of the full
non-linear relationship). In biological systems, this can be probed by
a broad-band noise stimulus. The Laplace transform of the linear
response function is known as the \emph{transfer function}, $H(s)$. It can be derived from the noisy phase oscillator
[Eq.~\eqref{eq.noisy_phaseoscil}], with PRC $Z(\phi)=\sum_k c_ke^{\imag 2\pi
k\phi}$, as \cite{schleimer_coding_2009}
\begin{equation} \label{eq.lrf}
  H(s)=\sum_{k=-\infty}^\infty\frac{s c_k}{s-\nu_k},
\end{equation}

with $s\in\mathbb C$ and poles at $\nu_k=-\imag2\pi k\frate-(\sigma
k)^2/2$ for $k \in \mathbb{Z}$. The transfer function $H(s)$ in Eq.~\eqref{eq.lrf}
establishes a link between PRC coefficients and filter characteristics.
The low frequency limit of the transfer spectrum, $|H(\imag2\pi f)|$, is
given by the mean for the PRC, $\lim_{f\to0}|H(\imag2\pi
f)|=c_0=\overline{Z(\phi)}$, and informs us whether the system can
track slow signals.  On the other hand, the higher the frequencies
present in $|H(\imag2\pi f)|$, the faster the signals a neuron can
follow.  Each non-zero Fourier mode of $Z(\phi)$ results in a
spectral peak.  The peaks of the spectrum are given by the poles
$\nu_k$, located at multiples of the neuron's mean spike rate,
$\frate$.

Near the SNL bifurcations, $H(s)$ passes considerably higher
frequencies than at the SNIC bifurcation [Fig.~\ref{fig:figure1}(c)],
because the number of spectral peaks is directly determined by the
higher number of Fourier modes in $Z_\text{SNL}$ (which is infinite 
in the case of the small SNL bifurcation).

Note that $Z_\text{Hopf}$ implies that the transfer spectrum of a
neuron near a supercritical Hopf bifurcation
has a single peak and suppresses low frequencies,
$|H(0)|=0$.  This is in contrast to the SNL, SNIC and HOM bifurcations,
which all pass arbitrarily slow signals. Although the
spectrum of the Hopf bifurcation has more power beyond its peak
compared to the SNIC neuron, it is only at the SNL and HOM bifurcations
that individual mean-driven neurons can truly follow frequencies much higher
than their own firing rate.

\subsection{Mutual information peaks around SNL points}\label{sec.info} 

According to the data processing inequality
\cite{cover_elements_2012}, the output of a neuron is an incomplete
representation of the information it receives as input. How much a
certain ensemble of input patterns, $I(t)$, is able to change a
neuron's output ensemble, $y(t)$, is quantified by the mutual
information rate, $M(y,I)$. A lower bound to the mutual information
rate, $M_\text{lb}\leqslant M$, results from the transfer 
function \cite{chacron_noise_2004, schleimer_coding_2009}

\begin{equation} \label{eq.info}
  M_\text{lb} =\int_0^{f_c}\mathrm df\ln
  \left(1+
\frac{|H(\imag2\pi f)|^2P_{II}(f)}{P_{y_0y_0}(f)}
\right),
\end{equation}

where $H$ is defined in Eq.~\eqref{eq.lrf}, and $P$ is the power
spectrum of input $I(t)$, respectively unperturbed output $y_0$. The
input spectrum $P_{II}$ is taken to be band-limited white noise with
cutoff frequency $f_c$, $P_{II}=\varepsilon^2/f_c,\; \forall
0\leqslant f\leqslant f_c$ and $P_{II}=0$ else.  According to
Eq.~\eqref{eq.info}, the information at a particular frequency is high
if the power of the output spectrum, $|H(\imag2\pi f)|^2P_{II}(f)$,
is high compared to the power of spike trains without input, $P_{y_0y_0}$.
$M_\mathrm{lb}$ sums up information in all frequency bands. The
increase of information transmission around the SNL point [Fig.~\ref{fig:figure2}(c)]
is a result of the facilitation of high
frequency transmission, \textit{cf.} Sec.~\ref{sec.lrf}.

\subsection{Spike-time reliability and locking to stimuli peak around SNL points}\label{sec.reliability}

The information transmission of a neuron also depends on how reliable
it encodes a time-varying stimulus into a sequence of spike times
\cite{mainen_reliability_1995}.  This can be quantified by observing
how fast an uncoupled population of identical neurons gets synchronized
by a common time-varying input (\emph{stochastic synchronization}
\cite{pikovsky_synchronization:_2003,marella_class-ii_2008}). Assuming
a set of identical phase oscillators from
Eq.~\eqref{eq.noisy_phaseoscil} with random initial conditions and
white noise input, the relaxation time constant is given as the
reciprocal of the Lyapunov exponent of the phase fixed points
\footnote{Assuming the phase difference $\psi=\phi_1-\phi_2$ of two
  phase oscillators (Eq.~\eqref{eq.noisy_phaseoscil} with common input
  $I(t)$ and individual noise $\xi(t)$) evolves slower than
  $\phi_{1,2}$ itself, averaging yields a Fokker-Planck equation for
  the phase difference: $\dot p(\psi)=\frac12\sigma^2p''(\psi)
  -\lambda(\psi p'(\psi)+p(\psi))$. The linear coefficient can be
  calculated by the Novikov-Furutzu-Donsker formula
  $\lambda=<Z'(\phi_2)x(t)>=-\varepsilon^2\int_0^1\text
  d\phi\,(Z'(\phi))^2$
  \cite{teramae_robustness_2004,goldobin_antireliability_2006}.  The
  steady density of the phase difference has variance
  $\lambda/\sigma^2$, which is related to the inverse of a spike
  metric.  All other eigenfunctions are Hermitian polynomials that
  decay with $(k\lambda)^{-1}$ \cite{gardiner_handbook_2004}. Both the
  decay to the stationary density and its variance are influence by the
  derivative of the PRC \cite{marella_class-ii_2008}.}
\begin{equation}\label{eq.lyap} \lambda=-\varepsilon^2\int_0^1
\left(\frac{\text{d}Z(\phi)}{\text{d}\phi}\right)^2\,\text{d}\phi.
\end{equation}
For neural computations, $\lambda^{-1}$ sets for example the
integration time scale a neuron requires to detect new transient
stimuli (\emph{evoked responses}). The interaction time scale is
minimized at the SNL bifurcations [Fig.~\ref{fig:figure2}(e)].  The
peaks of the Lyapunov exponent $\lambda$ furthermore imply efficient
locking to a common input.  High Fourier
modes, responsible for the peaks around the SNL points, are emphasized
by the derivative in Eq.~\eqref{eq.lyap}, which amounts to a
multiplication of the $k^{\text{th}}$-mode by $k$.  This results in an
even stronger stochastic synchronization than in the bi-phasic PRCs
emerging from some type-II excitable neurons
\cite{marella_class-ii_2008}.

\subsection{Synchronization peaks around SNL points}\label{sec.sync}

The asymmetry of the PRC scales the frequency detuning over which a
neuron entrains to its input (the width of the Arnold tongue
\cite{pikovsky_synchronization:_2003,kuramoto_chemical_1984}). The
input can either be a periodic signal or the recurrent input from other
neurons in a network. 
Here, we use synchronization in the sense
of a constant phase relation between oscillators, compare
Fig.~\ref{fig:syncDemo}.
The relation between PRC and synchronization can
be illustrated by two delta-coupled phase oscillators, $\phi_{1,2}$, as
defined in Eq.~\eqref{eq.noisy_phaseoscil},
\begin{equation}
  \dot\phi_{1,2}=f_{1,2} + Q(\phi_{1,2}-\phi_{2,1}) +
\sigma\,\xi_{1,2},
  \label{eq.phaseDetuning}
\end{equation}
where the coupling function $Q$ results from an averaging step if the
interaction between both oscillators are assumed to be weak
\cite{ermentrout_mathematical_2010}, $Q(\Delta) = \int_0^\infty
Z(\varphi)G(\varphi+\Delta) d\varphi\,$, where $G(\phi)$ is the time-varying
synaptic input evaluated on the limit cycle.  The phase difference,
$\psi=\phi_1-\phi_2$, evolves as $\dot\psi=\Delta f +
Q_\text{odd}(\psi)$, where $Q_\text{odd}(\psi)=Q(\psi)-Q(-\psi)$ is twice the
odd part of the coupling function.  Synchronization (\textit{i.e.}, a
constant phase lag $\psi$) requires $\dot\psi=0$, and the maximal
admissible frequency detuning $\Delta f$ is given by the image of
$Q_\text{odd}$.  In the case of $\delta$-coupling, $Q_\text{odd}$ is
equal to twice the odd part of the PRC, $Z_\text{odd}$, so that 
phase locking only occurs if
$\Delta f\in[\min Z_\text{odd}, \max Z_\text{odd}]$.
In Fig.~\ref{fig:figure2}(d), the synchronization range
$\max Z_\text{odd}-\max Z_\text{odd}$ is plotted. 
The increased synchronization range will also manifest itself in 
globally coupled networks of the type studied in 
Refs.~\cite{daido_order_1992, daido_onset_1996}.

The synchronization boost around the SNL points
[Fig.~\ref{fig:figure2}(d)] arises from period scaling and PRC asymmetry,
alongside a significant odd part in $Z_\text{SNL}$ compared to
$Z_\text{SNIC}$.  For two coupled oscillators, a small SNL bifurcation
favors alternated spiking, which is sometimes called
\emph{anti-synchronization}.  This is in contrast to the stable
in-phase locking that is observed for PRCs shaped like a negative sine
\cite[see appendix]{kirst_fundamental_2015}.

\subsection{Coding at an SNL bifurcation}

In summary, Secs.~\ref{sec.lrf} to \ref{sec.sync} suggest that passing
the SNL point affects a variety of neural coding schemes.
Regarding spike-timing codes, the shape of the PRC at the SNL
bifurcation is similar to the optimal PRC for synchronization by
common Poisson input \cite{hata_optimal_2011}. Furthermore, the lower
bound of the information rate is increased, because higher frequencies
emerge in the linear response function.  Regarding phase codes
\cite{friedrich_multiplexing_2004}, a consequence of the PRC at the
SNL bifurcation is a temporally more precise locking of spikes to a
reference signal. This prevents spurious spikes, \emph{i.e.}, fewer phase
slips occur, reducing the decoding error.  Regarding rate codes, the
lower bound of the decoding error (Cramer-Rao bound) is connected to
the slope of the firing rate as a function of the input $I_\text{DC}$
\cite{dayan_theoretical_2001}. At firing onset, this slope is known to
be larger for a HOM bifurcation compared to a SNIC bifurcation
\cite{izhikevich_dynamical_2007}, and hence rises around the SNL
point, potentially minimizing the decoding error.

\section{Generic occurrence of SNL bifurcations}\label{sec.generic}

The multifarious consequences of the SNL bifurcation discussed in
Sec.~\ref{sec.consequences} will be of particular relevance for
neuronal processing if the SNL bifurcation generally occurs in
realistic neuron models. Next, we demonstrate that indeed any 
two-dimensional, type-I conductance-based neuron model can always be
tuned to SNL bifurcations. More precisely, we show that the SNL
bifurcation is an essential element in the bifurcation diagram that
uses input current and membrane capacitance as control parameters. This
bifurcation diagram also allows us to relate the SNL bifurcation to
other bifurcations such as the Bogdanov-Takens (BT) bifurcation, 
classical termed the switch of type-I/II excitability
\cite{ermentrout_mathematical_2010,franci_balance_2013}.

    \begin{figure}
    \includegraphics[width=\columnwidth]{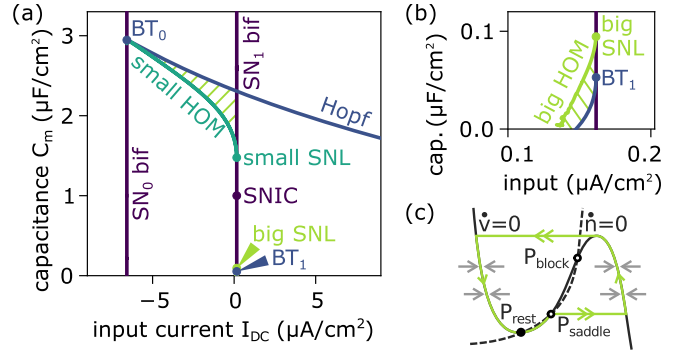}
    \caption{\label{fig:figure3}
    {Color) (a) Bifurcation diagram of the Wang-Buzsaki model under
variation of membrane capacitance $C_\text{m}$ and input current
$I_\text{DC}$.  With $C_\text{m}=1$ \si{\micro\farad}/cm$^2$, the limit cycle arises from a
SNIC bifurcation. Increasing $C_\text{m}$ leads to the small SNL
at $C_\text{m}\approx1.47$ \si{\micro\farad}/cm$^2$.
Dashed areas mark bistability. (b) Decreasing $C_\text{m}$ leads to
the big SNL, and then to a Bogdanov-Takens (BT) bifurcation. 
Note that a change of stability in the 
big HOM branch, not shown here, follows from Ref.~\cite{de_maesschalck_neural_2015}. 
(c) Schematic illustration for the
limit $C_\text{m}\to0$, in which the system corresponds to a
relaxation oscillator: Drawn in the state space of gating variable $n$
versus voltage $v$, the solid line with inverted N-shape represents
the voltage nullcline, and the dashed line represents the gating
nullcline. At some $I_\text{DC}$, the resting state looses stability
and a big HOM orbit around all fixed points (green) is created. }}
    \end{figure}

Concentrating on bifurcations relevant for neuronal spiking (\emph{i.e.},
bifurcations affecting a stable limit cycle), Fig.~\ref{fig:figure3}
shows a bifurcation diagram of the Wang-Buzsaki model
[Sec.~\ref{sec.modelDef}] with input
current and membrane capacitance as control parameters. Along the
dimension spanned by the capacitance, two SNL bifurcations enclose the
SNIC bifurcation. The lower SNL bifurcation corresponds to a big SNL
bifurcation for which the arising limit cycle encircles the ghost of
the saddle-node, and the
upper SNL bifurcation corresponds to a small SNL bifurcation for which
the ghost of the saddle-node lies outside of the limit cycle 
[Fig.~\ref{fig:figure1}(a)].  In particular, decreasing the
capacitance, an SNL point is passed \emph{before} the BT 
bifurcation is reached.

We show in Appendix~\ref{sec.proof} that this bifurcation structure
generalizes (under mild assumptions) to planar neuron models. The
membrane capacitance $C_{\text{m}}$ is used as bifurcation parameter
in the general bifurcation diagram that we construe, because it simply
changes the time scale of the voltage dynamics [Eq.~\eqref{eq:dynSys}].
The proof separately considers the lower and the upper part of the
bifurcation diagram. The lower part is based on the so-called
\emph{relaxation limit} with infinitely fast voltage dynamics that
arises from the limit $C_{\text{m}} \to 0$ [Fig.~\ref{fig:figure3}(c)], where the bifurcation
structure is known \cite{de_maesschalck_neural_2015}. Non-zero
capacitance values are deduced from several observations that restrict
the path of limit cycle bifurcation branches in planar systems. The
upper part of the bifurcation diagram is extracted from the unfolding
of a BT point.

Our derivation may be an interesting starting point for similar
results in other dynamical systems in which the time scale of a single
dynamical variable is used as bifurcation parameter. For our planar
neuron models, we find that the SNIC bifurcation branch is generically
enclosed by two SNL bifurcations that are reached by an adaptation of
the voltage time scale. In particular, our results show that a
continuous variation of the voltage time scale reaches the BT point
only \emph{after} passing one of the SNL bifurcations.

\section{Discussion}\label{sec.conclusion} 

This article explores the intricate relation between SNL bifurcations
and neuronal processing. SNL bifurcations entail optimal
synchronization and coding properties, with several measures of
information and synchronization peaking in the vicinity of SNL
bifurcations [Fig.~\ref{fig:figure2}].  Moreover, the approach to the
SNL point could be a mechanism to unify recent explanations of how
neurons transmit frequencies far above the cutoff given by their
membrane time constant \cite{naundorf_unique_2006, ilin_fast_2013,
ostojic_neuronal_2015, fourcaud_dynamics_2002}.  Indeed, homoclinic
neurons at the small SNL point seem to resemble the idealized
\emph{perfect integrate-and-fire} neuron
\cite{stein1972j:coherenceInfo}, as both are transmitting arbitrarily
high frequencies.

Drastic changes in spike-based coding can be expected at a bifurcation
that affects not only the fixed points, but also the stable limit
cycle. This is the case for the SNL bifurcation, in distinction to the
BT point classically referred to as the transition between type-I to
type-II excitability: In neuron models, the BT-associated Hopf
bifurcation is typically subcritical. The limit cycle arising at the
subcritical Hopf bifurcation is unstable and has only indirect
implications for spiking dynamics. The subcritical Hopf bifurcation
affects the resting state, changing subthreshold dynamics and filtering
\cite{richardson_subthreshold_2003,
schreiber_subthreshold_2004,ermentrout_mathematical_2010,
kirst_fundamental_2015}.  Spike-based coding is in this case only
affected if the system behaves like a fluctuation-driven escape problem
\cite{chow_spontaneous_1996,hong_single_2007}, while we focus on
mean-driven limit cycle dynamics.

Note that models in vicinity of a BT point have a different bifurcation
structure than models such as the original Hodgkin-Huxley (HH) model
\cite{hodgkin_quantitative_1952}.  While in the latter, the unstable
limit cycle that terminates in the subcritical Hopf bifurcation is born
at a fold of limit cycles bifurcation, the normal form of the BT
bifurcation shows that the unstable limit cycle is in the former case
born at a HOM bifurcation \footnote{The bifurcation structure of the HH
  model arises in our model when the two fold bifurcations collide in a
  cusp bifurcation. This can be achieved by parameters that affect the
shape of the nullclines, but is not possible with input and capacitance
as bifurcation parameters as used here.}. This difference will
probably affect the PRC of the stable limit cycle, for which the
canonical shape is still unresolved \footnote{The normal
  form used to calculate PRCs for the generalized Hopf bifurcation in
Ref.~ \cite{brown_phase_2004} assumes a circular symmetric fold of limit
cycles bifurcation.  This holds locally around the subcritical Hopf
bifurcation, but is unrealistic for full-blown pulse-like spikes, for
which a separation of time scales is required in the dynamics.}.
Moreover, the identified generic bifurcation sequence 
shows that a smooth change in time-scale parameters does not
justify the previously used heuristic formula that exploits a single Fourier mode
to interpolate between $Z_\text{SNIC}$ and $Z_\text{Hopf}$
\cite{tsubo_synchronization_2007, schleimer_coding_2009,
abouzeid_correlation_2011}. 

As codimension-two bifurcation, the SNL bifurcation is reached in
neuron models by an appropriate tuning of both the input current and
one additional model parameter. Examples for the second control
parameter are the membrane capacitance, maximal gating conductances,
tonic inhibition \cite{kirst_fundamental_2015}, neuromodulators
\cite{stiefel_effects_2009}, or gating time constants
\cite{izhikevich_dynamical_2007}. With the membrane capacitance as
bifurcation parameter, we demonstrate for planar conductance-based
models with a SNIC bifurcation that, ubiquitously, an SNL bifurcation
is the first bifurcation reached for lowered or increased capacitance,
respectively.  With the three bifurcation parameters capacitance,
input, and leak conductance, the identified sequence of bifurcations
collapses into a codimension-three cusp BT point
\cite{kirst_fundamental_2015, pereira_bogdanovtakens_2015}. This
potentially generalizes the described bifurcation structure beyond the
planar case. 

Focusing on neuron models that spike at low firing rates, where the
dynamics are dominated by the bifurcation that creates the limit cycle,
allows us to draw model-independent conclusions. Furthermore, the used
phase description demands for small inputs compared to the limit cycle
stability. The strong stability of the limit cycle around the SNL
point, shown by the Floquet exponent [Fig.~\ref{fig:figure2}(b)], validates
the phase reduction even for reasonably sized inputs.  Our setting with
low firing rates and relatively weak synaptic connections is typical
for cortical neurons \cite{barth_experimental_2012,
hoppensteadt_weakly_1997}.

Our mathematical arguments rely on the relative time scale between
state variables as the bifurcation parameter that takes us to the
relaxation limit used in Appendix \ref{sec.proof}.  The
membrane capacitance is one such parameter.  From a biological
perspective, the membrane capacitance is in several aspects an
interesting bifurcation parameter. The effective membrane capacitance
depends on cell parameters such as the morphology of the neuron or the
myelination of its axon \cite{hesse_externalization_2015}, and it may
hence be adapted on developmental or evolutionary time scales.
Furthermore, the experimental method of infrared neural stimulation
uses a change in membrane capacitance to depolarize neurons \cite{shapiro_infrared_2012}. The
neurons are stimulated by infrared laser pulses
\cite{wells_optical_2005}, and the deposition of energy leads to an
increase in the membrane capacitance,
changing the capacitive current $I_{\text{cap}} = \frac{\text{d}
C_\text{m} v}{\text{d} t} = \dot C_{\rm m}v + C_{\rm m} \dot v$. The
transient capacitance increase leads to a non-zero capacitive term
$\dot C_{\rm m}v$, which can be sufficient to excite the neuron
\cite{shapiro_infrared_2012}. Our study concentrates on neuronal
properties beyond this transient dynamics. We have demonstrated a rich
impact on neuronal dynamics of the specific value of $\Cm$, even in its
steady state with $\dot C_\text{m}=0$. For infrared neural stimulation,
this implies that the observed changes in capacitance, in addition to a
short-term excitation, could also push the neurons closer to an SNL
bifurcation, with potentially drastic implications for coding
properties. Moreover, passing the SNL point causes a region of
bistability between limit cycle and resting state $P_\text{rest}$
[Fig.~\ref{fig:figure3}(a-b)].  The consequential hysteresis results in
spiking that persists beyond the duration of the transient current
stimulation.  The respective switching statistics (\emph{i.e.}, between silent
periods and periodic firing) and their dependence on the timing of
excitation and inhibition may hence be characteristic of the different
SNL types and allow for a distinction between these dynamical regimes,
see also \cite{izhikevich_dynamical_2007}.  Overall, our results demand
for a careful interpretation of infrared neural stimulation experiments
because of our prediction that infrared neuronal stimulation may 
drive neurons into a significantly different dynamical regime.

Last but not least, our results suggest that also under natural
conditions, neurons may favor a position close to an SNL point in
order to profit from its unique coding flexibility. On the other hand,
SNL bifurcations may also be relevant in pathology (\emph{e.g.},
epilepsy).  Experimentally, the measurement of phase-response curves
is technically involved, even more so if multiple dynamical regimes
are probed in the same neuron.  Experimental reports exist for the
transition between type-I and type-II excitability
\cite{kirst_fundamental_2015, stiefel_effects_2009}. While several
studies report PRCs that are compatible with a spike onset either at a
SNIC bifurcation or at a HOM bifurcation \cite{schultheiss_phase_2011,
wang_hippocampal_2013, gutkin_phase-response_2005,
ermentrout_variance_2011}, the observation of an SNL bifurcation in an
experimental setting remains an open challenge. 

In summary, our study consists of two parts: We have first extracted
from the dynamics at an SNL bifurcation the phase-response curve, and
used this knowledge to infer the associated coding properties. Both the
PRC asymmetry and its high Fourier modes are generic properties at SNL
bifurcations. Thereby, our results are not only independent of
the particular neuron model, but are equally applicable to any system
that allows for a phase reduction. Moreover, we have demonstrated that
SNL bifurcations ubiquitously occur in a set of planar neuron models.
With the time scale of one dynamical variable as bifurcation parameter,
the structure of our proof is likely to generalize to other systems
with a subcritical Hopf bifurcation in the relaxation limit, such as
lasers \cite{eguia_semiconductor_1999,
martinez_avila_experimental_2004}, Josephson junctions
\cite{eksioglu_dissipative_2013, labouvie_bistability_2016, shimizu_relaxation_1995, romanelli_excitable-like_2016}, and chemical
reactions \cite{schneider_chemical_1991, balakotaiah_bifurcation_1999}.
Together, both parts highlight the SNL bifurcation as a hitherto
underestimated bifurcation with prominent importance for neuronal
dynamics.

\begin{acknowledgments} We thank Dr. Martin Wechselberger for inspiring
  discussions and pointing out the connection to his work on the
  relaxation limit and Grigory Bordyugov for advice on numerical
  continuation and homoclinic bifurcations. Funded by the German
Federal Ministry of Education and Research (01GQ0901, 01GQ1403).
\end{acknowledgments}

\appendix   

\section{Generic definition of conductance-based neuron models} 
\label{sec.modelDef}

We consider a generic class of conductance-based neuron models \cite{kirst_fundamental_2015}:

\begin{eqnarray}
\label{eq:condModel}
\dot{v}=  \frac{I_{\text{cap}}(v, ...)}{C_\text{m}}  &=&  \frac{1}{C_\text{m}}
   (I_{\text{in}}- g_{\text{L}}(v - v_{\text{L}}) - I_{\text{gating}}),\\ 
   I_{\text{gating}} &=& \sum_{i=0}^{n} g_{i} (v - v_i) \prod_{k=0}^K m_{ik}^{p_{ik}} 
\end{eqnarray}

where ion channel $i$ has maximal conductance $g_{i}$ and reversal
potential $v_i$ and its open probability is given by a product of
gating variables (potentially to some power of $p_{ik}$). Each gating
variable $m_{ik}$ of ion channel $i$ is either a function of the
voltage, $m_{ik} = m_{ik\infty}(v)$, or relaxes exponentially to its
steady state value $m_{ik\infty}(v)$, with gating kinetics given by

\begin{equation}
\dot{m_{ik}} = \frac{m_{ik\infty}(v)-m_{ik}}{\tau_{ik}(v)}
\end{equation}

For numerical continuation, we use a single-compartmental version of
the Wang-Buzsaki model for hippocampal pyramidal cells
\cite{wang_gamma_1996} with the following dynamics:

\begin{eqnarray*}
    \dot{v} &=& (I + g_{\text{L}} \, (E_{\text{L}}-v) + I_{\text{gate}})/C_{\text{m}}, \\ 
    \dot{h} &=& 5\;(\alpha_{h}(v) \, (1-h) - \beta_{h}(v) \, h), \\
    \dot{n} &=& 5\;(\alpha_{n}(v) \, (1-n) - \beta_{n}(v) \, n), 
\end{eqnarray*}
with membrane capacitance $C_{\text{m}}=1\mu \text{F/cm}^2$, 
maximal conductances
$g_{\text{L}}=0.1\mu \text{S/cm}^2$, $g_{\text{Na}}=35\mu
\text{S/cm}^2$,
$g_{\text{K}}=9\mu \text{S/cm}^2$, 
reversal potentials
$E_{\text{L}}=-65\text{mV}$, $E_{\text{Na}}=55\text{mV}$,
$E_{\text{K}}=-90\text{mV}$,
and the following functions:
\begin{eqnarray*}
 I_{\text{gate}} &=& g_{\text{Na}} 
m_\infty(v)^3   h (E_{\text{Na}}-v) + g_{\text{K}}  n^4 
(E_{\text{K}}-v), \\
    m_\infty &=& \frac{\frac{v+35}{ \exp(-0.1\,(v+35)) - 1 } }{
\frac{v+35}{ \exp(-0.1\,(v+35)) - 1 } - 40\, e^{-(v+60)/18}},\\
    \alpha_{h}(v) &=& 0.07 \, \exp(-(v+58)/20),\\
    \beta_{h}(v) &=& 1/( 1 + \exp(-0.1\,(v+28)) ),\\
    \alpha_{n}(v) &=& -0.01\,\frac{v+34}{\exp(-0.1\,(v+34)) - 1 },\\
    \beta_{n}(v) &=& 0.125\, \exp(-(v+44)/80).
\end{eqnarray*}

\section{PRC symmetry} \label{sec.prcsymAppendix}

The PRC asymmetry at the SNL bifurcation is a direct consequence of the
broken symmetry in the dynamics at the SNL bifurcation. This section
gives more detail on the relationship between dynamics and PRC, both
intuitively and with a mathematical argument. We will describe first
how the dynamics at the SNIC bifurcation leads to a symmetric PRC and
then show that these conditions are not met at the SNL bifurcation,
predicting an asymmetric PRC at an SNL bifurcation.  While the
arguments are presented with a small SNL bifurcation in mind, they hold
in a similar way for a big SNL bifurcation.

As introduced in the main text, the orbit at a SNIC bifurcation follows
the semi-stable manifold of the saddle-node fixed point, which
corresponds to the central manifold of a fold bifurcation. The zero
eigenvalue of the Jacobian $J$ at the saddle-node on the semi-stable
manifold eliminates the linear term. The leading second order term
results in a parabolic normal form. For dynamics centered around $x=0$,
stimulated with input $s$, the dynamics is

\begin{equation}
\dot{x}=s+x^2,
\label{eq:SNnormalForm}
\end{equation}

where all variables are chosen unitless for convenience. 

The dynamics is symmetric around the saddle-node fixed point, i.e.,
the orbit has corresponding velocities at approach and exit of the
saddle-node. The orbit flip at the SNL bifurcation breaks this
symmetry in the dynamics, and, as we will show, also in the PRC.

From a mathematical perspective, the normal form allows for a
calculation of the PRC. We, however, will use the normal form to
directly analyze PRC symmetry. For the SNIC bifurcation, the
reflection symmetry of the PRC can be inferred from the symmetry of the dynamics:
If $x(t)$ is a solution
of the dynamical system given by Eq.~\eqref{eq:SNnormalForm}, then the
same holds f or $-x(-t)$, $x(t)$ is hence point symmetric in time, $x(t)=-x(-t)$.  Derivation of
the right-hand side of Eq.~\eqref{eq:SNnormalForm} by $x$ results in a
Jacobian linear in $x$, that is hence point symmetric in $x$, $J(x) = -J(-x)$. Inserting both into
the adjoint equation, Eq.~\eqref{eq.adjoint}, directly leads to a PRC
reflection symmetric in time, $Z_\text{SNIC}(t)=Z_\text{SNIC}(-t)$.
In contrast, the asymmetric dynamics at the SNL bifurcation lead to an
asymmetric PRC. 

Intuitively, on an orbit that connects to a saddle-node fixed point,
the dynamics becomes arbitrary slow at the fixed point. The limit cycle
shows the slowest dynamics in the same region in state space, in
proximity to the ghost of the former saddle-node. A perturbation that
propels the dynamics over the ghost of the saddle-node will therefore
maximally advance the next spike.  The maximum of the PRC is at the
phase value that corresponds to the the saddle-node. For a SNIC
bifurcation, the PRC maximum lies at $\phi=0.5$ because the symmetric
dynamics of a SNIC take equal time for the approach (from $\phi=0$ to
$\phi=0.5$) and the exit (from $\phi=0.5$ to $\phi=1$) of the
saddle-node. In comparison, the PRC maximum is shifted towards the left
at the SNL bifurcation, because the accelerated entry along the
strongly-stable manifold advances the saddle-node to earlier phases.
The shift of the maximum away from the center destroys the symmetric
shape of the PRC.

The symmetry breaking generalizes beyond the SNL bifurcation: A saddle
homoclinic orbit shows an asymmetric PRC \cite{brown_phase_2004}, if
the saddle has different stabilities along stable and unstable
manifold, and hence non-symmetric dynamics. In summary, we showed that
the symmetry breaking in the PRC is an immediate consequence of the
symmetry breaking in the dynamics that occurs as orbit flip at the SNL
bifurcation. Hence, the observed symmetry breaking in the PRC is a
general property of the SNL bifurcation.

\section{Mathematical argument for the generic occurrence of the SNL
bifurcation in planar models} \label{sec.proof}

We show in the following that, with a variation of the time-scale
parameter, such capacitance, in a broad set of planar conductance-based
models, a SNIC bifurcation is always enclosed by two SNL bifurcations,
and that a decrease in capacitance passes the big SNL bifurcation, and
only afterwards reaches the BT point. Beyond the BT point, a Hopf
bifurcation destabilizes the resting state before the fold bifurcation
occurs.

To this aim, we prove that the general structure of the bifurcation
diagram [Fig.~\ref{fig:figure3}] holds for any planar neuron model
that conforms with our assumptions [Sec.~\ref{sec.planarModel}].

\subsection{Model definition}\label{sec.planarModel} 

We consider a generic class of type-I planar conductance-based neuron
models. The single gating variable, $n$, commonly models the opening
and closing of a restorative current originating, say, from the
potassium ion channel. The dynamics is given by

\begin{equation}
  \left(\begin{array}{c} \dot{v} \\ \dot{n}\end{array}\right)
        = F(v,n) 
  = \left(\begin{array}{c}
    \frac{1}{C_\text{m}}( I_{\text{DC}} - I_{\text{ion}} )
   \\ \frac{n_{\infty}(v) - n}{\tau_n(v)}
        \end{array}\right).
\label{eq:dynSystem}
\end{equation}
with $I_\text{ion}(v,n)= g_{\text{L}}(v-v_{\text{L}}) + g_{\text{K}}
\,n\,(v-v_{\text{K}})  $, compare Eq.~\ref{eq:condModel}.

We chose the model such that it fulfills the following assumptions: 

\paragraph*{(A1)} The firing onset of the model occurs, for some
capacitance value $C_\text{SNIC}$ and a specific input current
$I_\text{DC} = I_\text{SN1}$ (the \emph{threshold} current), at a
non-degenerated SNIC bifurcation.

\paragraph*{(A2)} We demand that at the capacitance 
$C_\text{SNIC}$, the subthreshold dynamics for
$I_\text{DC}<I_\text{SN1}$ relax to a single stable fixed point, the
\emph{resting state}. We furthermore assume that with an increase in
input current, the limit cycle dynamics eventually terminates in a
bifurcation denoted \emph{excitation block}, after which the dynamics
relaxes again to a stable fixed point. This assumption prevents
diverging dynamics.  

\paragraph*{(A3)} The nullcline of the voltage has an inverted
\emph{N}-shape.  \paragraph*{(A4)} We require that $n_{\infty}(v)$
from Eq.~\ref{eq:dynSystem} is an increasing, positive, bounded, twice
differentiable function that becomes sufficiently flat in the limit
$v\to\pm \infty$, $\lim_{v\to \pm \infty} v\, \partial_v n_{\infty}(v)
= 0$. This assumption allows us to use results from 
Ref.~\cite{kirst_fundamental_2015}. 

All of these assumptions are fulfilled in common neuron models
with type-I excitability. 

\subsection{Construction of the bifurcation diagram}

The following proof establishes an ordering in a sequence of limit
cycle bifurcations, whereby a SNIC is enclosed by two SNL bifurcations.
The ordering is established by analyzing the relaxation limit as an
anchoring point.  We thereby capitalize on recent results from the
relaxation limit, $C_{\text{m}} \to 0$. As we will show, the ordering
that arises in this limit along $I_{\text{DC}}$ implies the same
ordering along $C_{\text{m}}$, mainly because limit cycle bifurcation
branches cannot cross in planar systems.

The limit cycle bifurcation branches that we consider lie in the
region with $I_\text{DC} \le I_\text{SN1}$, because, for neuronal
firing, the limit cycle creation has to happen before (\emph{i.e.}, at
lower $I_{\text{DC}}$) or at the fold bifurcation at which the resting
state is eliminated.  $C_\text{SNIC}$ separates the region
$I_\text{DC} \le I_\text{SN1}$ in a lower and upper subregion. Since
the occurrence of limit cycle bifurcations at $C_\text{SNIC}$ is
prevented by the requirement \emph{(A2)} that stable dynamics are
given by a unique fixed point, all limit cycle bifurcation branches
lie either in one or the other subregion.
In the proof, we start with the lower subregion,
and then consider the upper one.

\subsection{The lower part of the bifurcation diagram,
$C_\text{m}<C_\text{SNIC}$}

\paragraph*{\textbf{Observation 0} Vertical fold bifurcation
branches: Fixed point location depends on $I_\text{DC}$, but not on
$C_\text{m}$.} The nullclines of Eq.~\ref{eq:dynSystem} are given by
$I_{\text{DC}}-I_\text{ion}(v,n)=0$ and $n=n_{\infty}(v)$. The nullclines are
independent of $C_\text{m}$, and therefore also the location of the
fixed points, because the fixed points sit at intersections of the
nullclines. Hence the location of the fold bifurcations is also
independent of $C_\text{m}$, which ensures that the fold branches
[marked with SN in Fig.~\ref{fig:figure3}(a)] are
vertical in a bifurcation diagram of $C_\text{m}$ versus
$I_\text{DC}$.

Based on the inverted-N shape of the voltage nullcline and the
monotonous shape of the gating nullcline, we can infer the existence
of one to three fixed points. For the following discussion, we name
these fixed points, a visualization of our nomenclature is shown in
Fig.~\ref{fig:figure3}(c). The number and location of the fixed points
is set by the input current $I_\text{DC}$, which shifts the voltage
nullcline up or down in the state space. For low, i.e., subthreshold
$I_\text{DC}$, the model has a single, stable fixed point,
$P_\text{rest}$. With an increase in $I_\text{DC}$, the knee of the
voltage nullcline approaches the gating nullcline from below, and
results in a fold bifurcation at some $I_\text{DC} = I_\text{SN0}$.
The fold bifurcation creates a saddle, $P_\text{saddle}$, and a node,
$P_\text{block}$. Our assumptions ensure that $P_\text{block}$ is
unstable because (A2) requires that $P_\text{rest}$ is the only stable
fixed point at $C_\text{m}=C_\text{SNIC}$.  Increasing the input
current further leads to a second fold bifurcation at some
$I_\text{DC} = I_\text{SN1}$.  This fold bifurcation annihilates
$P_\text{rest}$ and $P_\text{saddle}$. Beyond the bifurcation,
$P_\text{block}$ remains as the only surviving fixed point. 

The saddle fixed point $P_\text{saddle}$ only exists between
$I_\text{SN0}$ and $I_\text {SN1}$. The association of HOM bifurcations
with saddles directly constrains their bifurcation branches to the
region $I_\text{SN0} \le I_\text{DC} \le I_\text {SN1}$. In an
analogue way, Hopf bifurcation branches are constrained by the
existence of the associated focus fixed point: The Hopf branch that
destabilize the resting state $P_\text{rest}$ is restricted to input
currents below $I_\text{SN1}$, and the other Hopf branch that changes
the stability of $P_\text{block}$ is restricted input currents above
$I_\text{SN0}$. Further constraints will be developed throughout the
following arguments. 

\paragraph*{\textbf{Observation 1} Starting points for the branches of
big HOM and neighboring Hopf bifurcation: Anchoring the bifurcation
diagram in the limit $C_\text{m}\to0$ yields
$I_\text{big\,HOM}<I_\text{Hopf}$.} In the limit $C_\text{m}\to0$, the
conductance-based model is transformed into a relaxation oscillator
with voltage as fast variable, as sketched in
Fig.~\ref{fig:figure3}(c) \cite{izhikevich_slowly_2003}. For this
limit, de Maesschalck and Wechselberger have identified the full
bifurcation structure for generic planar neuron models
\cite{de_maesschalck_neural_2015}. Their \emph{Theorem 2} demonstrates
for sufficiently small $C_\text{m}$ that an increase in $I_\text{DC}$
results for model neurons such as ours in a generic sequence of
bifurcations. Relevant for our consideration is the occurrence of a
big HOM bifurcation at input $I_\text{big\,HOM}$, and a subcritical
Hopf bifurcation that destabilizes $P_\text{rest}$ at $I_\text{Hopf}$.
Their full bifurcation structure ensures furthermore that neither the
big HOM branch nor the Hopf branch returns to the limit
$C_\text{m}\to0$, which is important to ensure the existence of a
codimension-two bifurcation at the other end.  They state an ordering
of the bifurcation currents,
$I_\text{big\,HOM}<I_\text{Hopf}<I_\text{SN1}$, which will be used in
the following to infer the same ordering at finite values of
$C_\text{m}$.

\paragraph*{\textbf{Lemma 1} HOM branches cannot "bend backwards": A
variation in $C_\text{m}$ generically breaks homoclinic orbits to
hyperbolic fixed points.} In order to constrain the location of HOM
bifurcations in subsequent paragraphs, we want to show that the
tracing of a HOM branch leads us always in one direction along the
input current (increasing or decreasing input). 
Equivalently, we can show that a HOM branch cannot "bend backwards"
along the input current dimension. This is the case if we show that
HOM branches cannot have "vertical parts": A HOM branch cannot align
with a parameter variation exclusively in $C_\text{m}$, because, as we
show with this lemma, a variation in $C_\text{m}$ generically breaks
the homoclinic orbit.

Homoclinic orbits arise when the trajectory of the unstable direction
of a fixed point connects to its stable direction, i.e., stable and
unstable manifold overlap. A parameter variation can separate stable
and unstable manifold from each other, allowing for the definition of
a distance. This distance is measured by the so-called separation
function, $\text{sep}$ [Fig.~\ref{fig:bifOverview}(b)]. For parameter
values that lie on the HOM branch, the separation function is zero,
$\text{sep}(C_\text{HOM})=0$, and it increases to some finite value,
$\text{sep}(C_\text{m})>0$, if a variation in the parameter breaks the
homoclinic orbit, i.e., leaves the HOM branch. This is analogue to a
non-zero value of the partial derivative of the separation function,
which is known as the \emph{Melnikov integral}, $M$ [for a derivation
in planar systems see for example Ref.~\cite{chicone_ordinary_2006},
leading to Equation (6.12) that we use in Eq.~\ref{eq.melnikov}].

A variation in $C_\text{m}$ breaks the homoclinic orbit if the
corresponding Melnikov integral evaluated on the homoclinic orbit is
non-zero \cite{homburg_homoclinic_2010}. The Melnikov integral with
respect to $C_\text{m}$ for a homoclinic orbit with flow $h(t)$ is 
\begin{eqnarray}
  M &=& \int_{-\infty}^{\infty}      K(t)
  \, F(h(t))\, \frac{\partial  F(h(t))}{\partial C_\text{m}}
  \,\text{d} t\\
   &=& -\int_{-\infty}^{\infty}      K(t)
  \, \frac{(I_{\text{DC}} - I_{\text{ion}}(h(t)))^2}{C_\text{m}^{3}}  \,\text{d} t,
\label{eq.melnikov}
\end{eqnarray}
where $K(t) = \exp\left(-\int_{0}^{t} \text{div} \,
F(h(s))\,\text{d}s\right)$. For our system, the Melnikov integral is
strictly positive, $0 < M$, because $(i)$ $K(t)$ is, as an exponential
function, strictly positive, $\forall t:\;0<K(t)$, and $(ii)$, because
we implicitly assume the existence of a homoclinic orbit, the difference of
ionic and injected currents cannot be zero at all times, hence $\exists
t:\;(I_{\text{DC}} - I_{\text{ion}}(h(t)))^2>0$. With that, the capacitance breaks the
homoclinic orbit and thus tracing a HOM branch along one
direction results either in continuously increasing or decreasing
input current values on the branch. This lemma is used in the
following Observation 2 in order to pursue the big HOM branch 
starting in the limit $\Cm \to 0$ [see Observation 1].

\paragraph*{\textbf{Observation 2} The big HOM branch eventually
approaches the fold bifurcation at $I_\text{SN1}$.} Based on the
directionality of the big HOM branch derived in the literature, we
will show in this observation that the big HOM branch eventually
approaches the fold bifurcation branch at which the resting state
collides with the saddle. The point of contact corresponds to an SNL
bifurcation, as we will show in subsequent paragraphs.

The statement of Theorem 2 by de Maesschalck and Wechselberger states
for sufficiently small $C_\text{m}$, in addition to the ordering used
in Observation 1, that the big HOM branch departs from its starting
point to the right, i.e., in the direction of increasing input current
\cite{de_maesschalck_neural_2015}. This directionality of the big HOM
branch generalizes to larger values of $\Cm$, because Lemma 1 prevents
"backward bends" of HOM branches. Given that the big HOM branch does
not return to the limit $C_\text{m}\to0$ [Observation 1], the big HOM
branch eventually has to approach the fold bifurcation at
$I_\text{SN1}$. The next lemma ensures that the connection point is an
SNL point.

\paragraph*{\textbf{Lemma 2} A HOM branch and the fold
branch at $I_\text{SN1}$ connect in an SNL bifurcation: A HOM branch is
stable when it connects to a non-degenerated fold bifurcation
involving a stable node.} An SNL bifurcation involves a stable
homoclinic orbit that transitions between a HOM
bifurcation and a SNIC bifurcation. The homoclinic orbit of the HOM
branch is stable, if the associated saddle-quantity is negative (the
sum of the two eigenvalues of the associated fixed point). At the
connection point with the fold branch, the homoclinic orbit is
associated with a saddle-node fixed point arising from the collision
of a stable node and a saddle. It has one zero eigenvalue (prerequisite
for the fold bifurcation) and one negative eigenvalue (the former
stable node sets the stability of the strongly-stable manifold). The
sum evaluates to a negative saddle-quantity, ensuring a stable
homoclinic orbit, and hence an SNL bifurcation.

\paragraph*{\textbf{Lemma 3} The bifurcation sequence in the lower
part of the bifurcation diagram: For $I_\text{DC} = I_\text{SN1}$,
increasing $C_\text{m}$ from zero passes first a BT point, then an SNL
point, before a non-degenerated SNIC bifurcation occurs,
$C_\text{BT1}<C_\text{big\,SNL}<C_\text{SNIC}$.} Combining Observation
2 and Lemma 2, we conclude that the big HOM branch connects to the
fold bifurcation branch at $I_\text{SN1}$ with a stable homoclinic
orbit, i.e., in an SNL bifurcation. This big SNL bifurcation happens
at some point $(I_\text{SN1}, C_\text{big\,SNL})$, with
$C_\text{big\,SNL}<C_\text{SNIC}$ because the big HOM branch cannot
pass the capacitance value of $C_\text{SNIC}$ as \emph{(A2)} prohibits
stable limit cycle bifurcations for $I_\text{DC}<I_\text{SN1}$.
From Observation 1, we know for $C_\text{m}\to0$ that a Hopf branch
starts at $I_\text{Hopf}$, and that this branch does not return to the
limit $C_\text{m}\to0$.  Because limit cycle bifurcation branches
cannot cross each other in a planar system, the Hopf branch can
furthermore not cross the big HOM branch. Instead, it connects to the
fold bifurcation branch in a BT bifurcation at some point
$(I_\text{SN1}, C_\text{BT1})$. The ordering
$I_\text{big\,HOM}<I_\text{Hopf}$ from Observation 1 immediately
implies an ordering in $C_\text{m}$, i.e.,
$C_\text{BT1}<C_\text{big\,SNL}$.  In summary, we have shown in this
lemma that $C_\text{BT1}<C_\text{big\,SNL}<C_\text{SNIC}$.

These arguments have proven the bifurcation sequence in the lower part
of the bifurcation diagram arising from the limit $C_\text{m}\to0$. In
the following, we use the unfolding of a second BT point to show the
upper part of the bifurcation diagram.

\subsection{The upper part of the bifurcation diagram,
$C_\text{m}>C_\text{SNIC}$}

\paragraph*{\textbf{Observation 3} The bifurcation diagram contains
exactly two BT points.} Kirst et al. identified the BT point for a
generic class of conductance-based neuron model (including our model
group) at a capacitance value that can be calculated from the input
current at which the fold bifurcation occurs
[Ref.~\cite{kirst_fundamental_2015}, Supplemental Material]. With the
two fold bifurcation branches occurring in our model group at input
currents $I_\text{SN0}$ and $I_\text{SN1}$, we find one unique BT
point on each fold branches. Lemma 3 identified one of them at the BT
point $(I_\text{SN1}, C_\text{BT1})$, and the second BT bifurcation
occurs at some point $(I_\text{SN0}, C_\text{BT0})$. From the BT point
at $(I_\text{SN0}, C_\text{BT0})$ arises by normal form theory a Hopf
bifurcation branch and a branch of a small HOM bifurcation. Both
depart in the direction of increasing input $I_\text{DC}$, which will
be used as before to constrain their location.

\paragraph*{\textbf{Observation 4} The second BT point lies
in the upper part of the bifurcation diagram: The BT point at
$(I_\text{SN0}, C_\text{BT0})$ occurs at $C_\text{BT0} >
C_\text{SNIC}$.} We restrict the region accessible to the Hopf branch
that arises from the BT point at $(I_\text{SN0}, C_\text{BT0})$: A
limit cycle bifurcation branch cannot cross other limit cycle
bifurcation branches (in a planar system), and hence the Hopf branch
cannot pass the SNIC bifurcation line between $(I_\text{SN1},
C_\text{SNIC})$ and $(I_\text{SN1}, C_\text{big\,SNL}) $, nor the big
HOM branch. Furthermore, (A2) demands that no stable fixed point
exists for $I_\text{DC} < I_\text{SN1}$ for $C_\text{m} =
C_\text{SNIC}$, effectively preventing the Hopf branch to pass this
line. The Hopf branch lies hence either entirely within or outside of
the region bounded by these lines.

We show that the Hopf branch lies outside of this region by
identifying this branch with the excitation block occurring at
$C_\text{m} = C_\text{SNIC}$: (A2) demands that the excitation block
at some $I_\text{DC} > I_\text{SN1}$, i.e., outside of the identified
region. Around the excitation block, $P_\text{block}$ is stabilized by
a Hopf bifurcation. This Hopf bifurcation affect $P_\text{block}$ and
hence belongs to the same branch of Hopf bifurcations that arises at
the BT point at $(I_\text{SN0}, C_\text{BT0})$, because this is where
$P_\text{block}$ is created. With that, the Hopf branch must lie
outside the region denoted above, and correspondingly also the BT
point at its end. We hence conclude $C_\text{BT0} > C_\text{SNIC}$.

\paragraph*{\textbf{Lemma 4} The small SNL bifurcation: A second SNL
bifurcation occurs at some $C_\text{small\,SNL}>C_\text{SNIC}$.} The
branch of the small HOM bifurcation that arises from the BT point at
$(I_\text{SN0}, C_\text{BT0})$ [see Observation 3] continues by Lemma
1 in the direction of increasing input $I_\text{DC}$. Hence, we find
some $C_\text{m} = C_\text{small\,SNL}$ for which the small HOM branch
connects to the fold bifurcation at $I_\text{SN1}$. At the connection
point, the HOM branch must be stable by Lemma 2. We identify the point
$(I_\text{SN1}, C_\text{small\,SNL})$ as small SNL bifurcation. 

For the overall proof, it remains to show the ordering
$C_\text{small\,SNL} > C_\text{SNIC}$. For that, we observe that a
limit cycle exists between the small HOM and the Hopf branch arising
from the BT point and contrast this with the limit cycle arising from
the SNIC bifurcation. As the Hopf bifurcation has to terminate the
limit cycle of the SNIC bifurcation at $C_\text{SNIC}$ [\emph{(A2)}],
it cannot terminate the limit cycle arising from the small HOM
bifurcation at this capacitance value.  This only leaves the
possibility for the SNL point to occur at some $C_\text{small\,SNL} >
C_\text{SNIC}$.

In summary, we have shown that $C_\text{BT1} < C_\text{big\,SNL} <
C_\text{SNIC} < C_\text{small\,SNL}$. This generic bifurcation
structure occurs with the membrane capacitance $C_\text{m}$ as
bifurcation parameter at $I_\text{DC} = I_\text{SN1}$. For a model
starting at a SNIC bifurcation, a variation in the capacitance will
thus pass an SNL bifurcation before a BT point is reached.

\bibliography{arxiv}

\begin{thebibliography}{95}%
\makeatletter
\providecommand \@ifxundefined [1]{%
 \@ifx{#1\undefined}
}%
\providecommand \@ifnum [1]{%
 \ifnum #1\expandafter \@firstoftwo
 \else \expandafter \@secondoftwo
 \fi
}%
\providecommand \@ifx [1]{%
 \ifx #1\expandafter \@firstoftwo
 \else \expandafter \@secondoftwo
 \fi
}%
\providecommand \natexlab [1]{#1}%
\providecommand \enquote  [1]{``#1''}%
\providecommand \bibnamefont  [1]{#1}%
\providecommand \bibfnamefont [1]{#1}%
\providecommand \citenamefont [1]{#1}%
\providecommand \href@noop [0]{\@secondoftwo}%
\providecommand \href [0]{\begingroup \@sanitize@url \@href}%
\providecommand \@href[1]{\@@startlink{#1}\@@href}%
\providecommand \@@href[1]{\endgroup#1\@@endlink}%
\providecommand \@sanitize@url [0]{\catcode `\\12\catcode `\$12\catcode
  `\&12\catcode `\#12\catcode `\^12\catcode `\_12\catcode `\%12\relax}%
\providecommand \@@startlink[1]{}%
\providecommand \@@endlink[0]{}%
\providecommand \url  [0]{\begingroup\@sanitize@url \@url }%
\providecommand \@url [1]{\endgroup\@href {#1}{\urlprefix }}%
\providecommand \urlprefix  [0]{URL }%
\providecommand \Eprint [0]{\href }%
\providecommand \doibase [0]{http://dx.doi.org/}%
\providecommand \selectlanguage [0]{\@gobble}%
\providecommand \bibinfo  [0]{\@secondoftwo}%
\providecommand \bibfield  [0]{\@secondoftwo}%
\providecommand \translation [1]{[#1]}%
\providecommand \BibitemOpen [0]{}%
\providecommand \bibitemStop [0]{}%
\providecommand \bibitemNoStop [0]{.\EOS\space}%
\providecommand \EOS [0]{\spacefactor3000\relax}%
\providecommand \BibitemShut  [1]{\csname bibitem#1\endcsname}%
\let\auto@bib@innerbib\@empty
\bibitem [{\citenamefont {Peron}\ and\ \citenamefont
  {Gabbiani}(2009)}]{peron_spike_2009}%
  \BibitemOpen
  \bibfield  {author} {\bibinfo {author} {\bibfnamefont {Simon}\ \bibnamefont
  {Peron}}\ and\ \bibinfo {author} {\bibfnamefont {Fabrizio}\ \bibnamefont
  {Gabbiani}},\ }\bibfield  {title} {\enquote {\bibinfo {title} {Spike
  frequency adaptation mediates looming stimulus selectivity in a
  collision-detecting neuron},}\ }\href {\doibase 10.1038/nn.2259} {\bibfield
  {journal} {\bibinfo  {journal} {Nature Neuroscience}\ }\textbf {\bibinfo
  {volume} {12}},\ \bibinfo {pages} {318--326} (\bibinfo {year}
  {2009})}\BibitemShut {NoStop}%
\bibitem [{\citenamefont {Laughlin}\ and\ \citenamefont
  {Sejnowski}(2003)}]{laughlin_communication_2003}%
  \BibitemOpen
  \bibfield  {author} {\bibinfo {author} {\bibfnamefont {Simon~B.}\
  \bibnamefont {Laughlin}}\ and\ \bibinfo {author} {\bibfnamefont
  {Terrence~J.}\ \bibnamefont {Sejnowski}},\ }\bibfield  {title} {\enquote
  {\bibinfo {title} {Communication in {Neuronal} {Networks}},}\ }\href
  {\doibase 10.1126/science.1089662} {\bibfield  {journal} {\bibinfo  {journal}
  {Science}\ }\textbf {\bibinfo {volume} {301}},\ \bibinfo {pages} {1870--1874}
  (\bibinfo {year} {2003})}\BibitemShut {NoStop}%
\bibitem [{\citenamefont {Niven}\ and\ \citenamefont
  {Laughlin}(2008)}]{niven_energy_2008}%
  \BibitemOpen
  \bibfield  {author} {\bibinfo {author} {\bibfnamefont {Jeremy~E.}\
  \bibnamefont {Niven}}\ and\ \bibinfo {author} {\bibfnamefont {Simon~B.}\
  \bibnamefont {Laughlin}},\ }\bibfield  {title} {\enquote {\bibinfo {title}
  {Energy limitation as a selective pressure on the evolution of sensory
  systems},}\ }\href {\doibase 10.1242/jeb.017574} {\bibfield  {journal}
  {\bibinfo  {journal} {Journal of Experimental Biology}\ }\textbf {\bibinfo
  {volume} {211}},\ \bibinfo {pages} {1792--1804} (\bibinfo {year}
  {2008})}\BibitemShut {NoStop}%
\bibitem [{\citenamefont {Pouget}\ \emph {et~al.}(2000)\citenamefont {Pouget},
  \citenamefont {Dayan},\ and\ \citenamefont
  {Zemel}}]{pouget_information_2000}%
  \BibitemOpen
  \bibfield  {author} {\bibinfo {author} {\bibfnamefont {Alexandre}\
  \bibnamefont {Pouget}}, \bibinfo {author} {\bibfnamefont {Peter}\
  \bibnamefont {Dayan}}, \ and\ \bibinfo {author} {\bibfnamefont {Richard}\
  \bibnamefont {Zemel}},\ }\bibfield  {title} {\enquote {\bibinfo {title}
  {Information processing with population codes},}\ }\href {\doibase
  10.1038/35039062} {\bibfield  {journal} {\bibinfo  {journal} {Nature Reviews
  Neuroscience}\ }\textbf {\bibinfo {volume} {1}},\ \bibinfo {pages} {125--132}
  (\bibinfo {year} {2000})}\BibitemShut {NoStop}%
\bibitem [{\citenamefont {Koch}\ and\ \citenamefont
  {Segev}(2000)}]{koch_role_2000}%
  \BibitemOpen
  \bibfield  {author} {\bibinfo {author} {\bibfnamefont {Christof}\
  \bibnamefont {Koch}}\ and\ \bibinfo {author} {\bibfnamefont {Idan}\
  \bibnamefont {Segev}},\ }\bibfield  {title} {\enquote {\bibinfo {title} {The
  role of single neurons in information processing},}\ }\href {\doibase
  10.1038/81444} {\bibfield  {journal} {\bibinfo  {journal} {Nature
  Neuroscience}\ }\textbf {\bibinfo {volume} {3}},\ \bibinfo {pages}
  {1171--1177} (\bibinfo {year} {2000})}\BibitemShut {NoStop}%
\bibitem [{\citenamefont {Schreiber}\ \emph {et~al.}(2002)\citenamefont
  {Schreiber}, \citenamefont {Machens}, \citenamefont {Herz},\ and\
  \citenamefont {Laughlin}}]{schreiber_energy-efficient_2002}%
  \BibitemOpen
  \bibfield  {author} {\bibinfo {author} {\bibfnamefont {Susanne}\ \bibnamefont
  {Schreiber}}, \bibinfo {author} {\bibfnamefont {Christian~K.}\ \bibnamefont
  {Machens}}, \bibinfo {author} {\bibfnamefont {Andreas. V.~M.}\ \bibnamefont
  {Herz}}, \ and\ \bibinfo {author} {\bibfnamefont {Simon~B.}\ \bibnamefont
  {Laughlin}},\ }\bibfield  {title} {\enquote {\bibinfo {title}
  {Energy-{Efficient} {Coding} with {Discrete} {Stochastic} {Events}},}\ }\href
  {\doibase 10.1162/089976602753712963} {\bibfield  {journal} {\bibinfo
  {journal} {Neural Computation}\ }\textbf {\bibinfo {volume} {14}},\ \bibinfo
  {pages} {1323--1346} (\bibinfo {year} {2002})}\BibitemShut {NoStop}%
\bibitem [{\citenamefont {Hesse}\ and\ \citenamefont
  {Schreiber}(2015)}]{hesse_externalization_2015}%
  \BibitemOpen
  \bibfield  {author} {\bibinfo {author} {\bibfnamefont {Janina}\ \bibnamefont
  {Hesse}}\ and\ \bibinfo {author} {\bibfnamefont {Susanne}\ \bibnamefont
  {Schreiber}},\ }\bibfield  {title} {\enquote {\bibinfo {title}
  {Externalization of neuronal somata as an evolutionary strategy for energy
  economization},}\ }\href {\doibase 10.1016/j.cub.2015.02.024} {\bibfield
  {journal} {\bibinfo  {journal} {Current Biology}\ }\textbf {\bibinfo {volume}
  {25}},\ \bibinfo {pages} {R324--R325} (\bibinfo {year} {2015})}\BibitemShut
  {NoStop}%
\bibitem [{\citenamefont {Sato}\ and\ \citenamefont
  {Aihara}(2014)}]{sato_changes_2014}%
  \BibitemOpen
  \bibfield  {author} {\bibinfo {author} {\bibfnamefont {Yasuomi~D.}\
  \bibnamefont {Sato}}\ and\ \bibinfo {author} {\bibfnamefont {Kazuyuki}\
  \bibnamefont {Aihara}},\ }\bibfield  {title} {\enquote {\bibinfo {title}
  {Changes of {Firing} {Rate} {Induced} by {Changes} of {Phase} {Response}
  {Curve} in {Bifurcation} {Transitions}},}\ }\href {\doibase
  10.1162/NECO_a_00653} {\bibfield  {journal} {\bibinfo  {journal} {Neural
  Computation}\ }\textbf {\bibinfo {volume} {26}},\ \bibinfo {pages}
  {2395--2418} (\bibinfo {year} {2014})}\BibitemShut {NoStop}%
\bibitem [{\citenamefont {Dayan}\ and\ \citenamefont
  {Abbott}(2001)}]{dayan_theoretical_2001}%
  \BibitemOpen
  \bibfield  {author} {\bibinfo {author} {\bibfnamefont {Peter}\ \bibnamefont
  {Dayan}}\ and\ \bibinfo {author} {\bibfnamefont {Larry~F.}\ \bibnamefont
  {Abbott}},\ }\href@noop {} {\emph {\bibinfo {title} {Theoretical
  {Neuroscience}: {Computational} and {Mathematical} {Modeling} of {Neural}
  {Systems}}}}\ (\bibinfo  {publisher} {Massachusetts Institute of Technology
  Press},\ \bibinfo {year} {2001})\BibitemShut {NoStop}%
\bibitem [{\citenamefont {Traub}\ \emph {et~al.}(1991)\citenamefont {Traub},
  \citenamefont {Wong}, \citenamefont {Miles},\ and\ \citenamefont
  {Michelson}}]{traub_model_1991}%
  \BibitemOpen
  \bibfield  {author} {\bibinfo {author} {\bibfnamefont {Roger~D.}\
  \bibnamefont {Traub}}, \bibinfo {author} {\bibfnamefont {Robert~K.}\
  \bibnamefont {Wong}}, \bibinfo {author} {\bibfnamefont {Richard}\
  \bibnamefont {Miles}}, \ and\ \bibinfo {author} {\bibfnamefont {Hillary}\
  \bibnamefont {Michelson}},\ }\bibfield  {title} {\enquote {\bibinfo {title}
  {A model of a {CA}3 hippocampal pyramidal neuron incorporating voltage-clamp
  data on intrinsic conductances},}\ }\href
  {http://jn.physiology.org/content/66/2/635} {\bibfield  {journal} {\bibinfo
  {journal} {{Journal of Neurophysiology}}\ }\textbf {\bibinfo {volume} {66}},\
  \bibinfo {pages} {635--650} (\bibinfo {year} {1991})}\BibitemShut {NoStop}%
\bibitem [{\citenamefont {Wang}\ and\ \citenamefont
  {Buzs{\'a}ki}(1996)}]{wang_gamma_1996}%
  \BibitemOpen
  \bibfield  {author} {\bibinfo {author} {\bibfnamefont {Xiao-Jing}\
  \bibnamefont {Wang}}\ and\ \bibinfo {author} {\bibfnamefont {Gy{\"o}rgy}\
  \bibnamefont {Buzs{\'a}ki}},\ }\bibfield  {title} {\enquote {\bibinfo {title}
  {Gamma {Oscillation} by {Synaptic} {Inhibition} in a {Hippocampal}
  {Interneuronal} {Network} {Model}},}\ }\href
  {http://www.jneurosci.org/content/16/20/6402} {\bibfield  {journal} {\bibinfo
   {journal} {The Journal of Neuroscience}\ }\textbf {\bibinfo {volume} {16}},\
  \bibinfo {pages} {6402--6413} (\bibinfo {year} {1996})}\BibitemShut {NoStop}%
\bibitem [{\citenamefont {Ilin}\ \emph {et~al.}(2013)\citenamefont {Ilin},
  \citenamefont {Malyshev}, \citenamefont {Wolf},\ and\ \citenamefont
  {Volgushev}}]{ilin_fast_2013}%
  \BibitemOpen
  \bibfield  {author} {\bibinfo {author} {\bibfnamefont {Vladimir}\
  \bibnamefont {Ilin}}, \bibinfo {author} {\bibfnamefont {Aleksey}\
  \bibnamefont {Malyshev}}, \bibinfo {author} {\bibfnamefont {Fred}\
  \bibnamefont {Wolf}}, \ and\ \bibinfo {author} {\bibfnamefont {Maxim}\
  \bibnamefont {Volgushev}},\ }\bibfield  {title} {\enquote {\bibinfo {title}
  {Fast {Computations} in {Cortical} {Ensembles} {Require} {Rapid} {Initiation}
  of {Action} {Potentials}},}\ }\href {\doibase 10.1523/JNEUROSCI.0771-12.2013}
  {\bibfield  {journal} {\bibinfo  {journal} {Journal of Neuroscience}\
  }\textbf {\bibinfo {volume} {33}},\ \bibinfo {pages} {2281--2292} (\bibinfo
  {year} {2013})}\BibitemShut {NoStop}%
\bibitem [{\citenamefont {Hong}\ \emph {et~al.}(2007)\citenamefont {Hong},
  \citenamefont {Ag\:{u}era~y Arcas},\ and\ \citenamefont
  {Fairhall}}]{hong_single_2007}%
  \BibitemOpen
  \bibfield  {author} {\bibinfo {author} {\bibfnamefont {Sungho}\ \bibnamefont
  {Hong}}, \bibinfo {author} {\bibfnamefont {Blaise}\ \bibnamefont
  {Ag\:{u}era~y Arcas}}, \ and\ \bibinfo {author} {\bibfnamefont {Adrienne~L.}\
  \bibnamefont {Fairhall}},\ }\bibfield  {title} {\enquote {\bibinfo {title}
  {Single {Neuron} {Computation}: {From} {Dynamical} {System} to {Feature}
  {Detector}},}\ }\href {\doibase 10.1162/neco.2007.19.12.3133} {\bibfield
  {journal} {\bibinfo  {journal} {Neural Computation}\ }\textbf {\bibinfo
  {volume} {19}},\ \bibinfo {pages} {3133--3172} (\bibinfo {year}
  {2007})}\BibitemShut {NoStop}%
\bibitem [{\citenamefont {Brunel}\ \emph {et~al.}(2003)\citenamefont {Brunel},
  \citenamefont {Hakim},\ and\ \citenamefont
  {Richardson}}]{brunel_firing-rate_2003}%
  \BibitemOpen
  \bibfield  {author} {\bibinfo {author} {\bibfnamefont {Nicolas}\ \bibnamefont
  {Brunel}}, \bibinfo {author} {\bibfnamefont {Vincent}\ \bibnamefont {Hakim}},
  \ and\ \bibinfo {author} {\bibfnamefont {Magnus J.~E.}\ \bibnamefont
  {Richardson}},\ }\bibfield  {title} {\enquote {\bibinfo {title} {Firing-rate
  resonance in a generalized integrate-and-fire neuron with subthreshold
  resonance},}\ }\href {\doibase 10.1103/PhysRevE.67.051916} {\bibfield
  {journal} {\bibinfo  {journal} {Physical Review E}\ }\textbf {\bibinfo
  {volume} {67}},\ \bibinfo {pages} {051916} (\bibinfo {year}
  {2003})}\BibitemShut {NoStop}%
\bibitem [{\citenamefont {Wei}\ and\ \citenamefont
  {Wolf}(2011)}]{wei_spike_2011}%
  \BibitemOpen
  \bibfield  {author} {\bibinfo {author} {\bibfnamefont {Wei}\ \bibnamefont
  {Wei}}\ and\ \bibinfo {author} {\bibfnamefont {Fred}\ \bibnamefont {Wolf}},\
  }\bibfield  {title} {\enquote {\bibinfo {title} {Spike {Onset} {Dynamics} and
  {Response} {Speed} in {Neuronal} {Populations}},}\ }\href {\doibase
  10.1103/PhysRevLett.106.088102} {\bibfield  {journal} {\bibinfo  {journal}
  {Physical Review Letters}\ }\textbf {\bibinfo {volume} {106}},\ \bibinfo
  {pages} {088102} (\bibinfo {year} {2011})}\BibitemShut {NoStop}%
\bibitem [{\citenamefont {Schleimer}\ and\ \citenamefont
  {Stemmler}(2009)}]{schleimer_coding_2009}%
  \BibitemOpen
  \bibfield  {author} {\bibinfo {author} {\bibfnamefont {Jan-Hendrik}\
  \bibnamefont {Schleimer}}\ and\ \bibinfo {author} {\bibfnamefont {Martin}\
  \bibnamefont {Stemmler}},\ }\bibfield  {title} {\enquote {\bibinfo {title}
  {Coding of {Information} in {Limit} {Cycle} {Oscillators}},}\ }\href
  {\doibase 10.1103/PhysRevLett.103.248105} {\bibfield  {journal} {\bibinfo
  {journal} {Physical Review Letters}\ }\textbf {\bibinfo {volume} {103}},\
  \bibinfo {pages} {248105} (\bibinfo {year} {2009})}\BibitemShut {NoStop}%
\bibitem [{\citenamefont {Hodgkin}(1948)}]{hodgkin_local_1948}%
  \BibitemOpen
  \bibfield  {author} {\bibinfo {author} {\bibfnamefont {Alan~L.}\ \bibnamefont
  {Hodgkin}},\ }\bibfield  {title} {\enquote {\bibinfo {title} {The local
  electric changes associated with repetitive action in a non-medullated
  axon},}\ }\href {http://www.ncbi.nlm.nih.gov/pmc/articles/PMC1392160/}
  {\bibfield  {journal} {\bibinfo  {journal} {The {Journal} of {Physiology}}\
  }\textbf {\bibinfo {volume} {107}},\ \bibinfo {pages} {165--181} (\bibinfo
  {year} {1948})}\BibitemShut {NoStop}%
\bibitem [{\citenamefont {Ermentrout}\ and\ \citenamefont
  {Kopell}(1986)}]{ermentrout_parabolic_1986}%
  \BibitemOpen
  \bibfield  {author} {\bibinfo {author} {\bibfnamefont {G.~Bard}\ \bibnamefont
  {Ermentrout}}\ and\ \bibinfo {author} {\bibfnamefont {Nancy}\ \bibnamefont
  {Kopell}},\ }\bibfield  {title} {\enquote {\bibinfo {title} {Parabolic
  bursting in an excitable system coupled with a slow oscillation},}\
  }\href@noop {} {\bibfield  {journal} {\bibinfo  {journal} {SIAM Journal on
  Applied Mathematics}\ }\textbf {\bibinfo {volume} {46}},\ \bibinfo {pages}
  {233--253} (\bibinfo {year} {1986})}\BibitemShut {NoStop}%
\bibitem [{\citenamefont {Rinzel}\ and\ \citenamefont
  {Ermentrout}(1998)}]{rinzel_analysis_1998}%
  \BibitemOpen
  \bibfield  {author} {\bibinfo {author} {\bibfnamefont {John}\ \bibnamefont
  {Rinzel}}\ and\ \bibinfo {author} {\bibfnamefont {G.~Bard}\ \bibnamefont
  {Ermentrout}},\ }\enquote {\bibinfo {title} {Analysis of neural excitability
  and oscillations},}\ in\ \href@noop {} {\emph {\bibinfo {booktitle} {Methods
  in neuronal modeling}}},\ \bibinfo {editor} {edited by\ \bibinfo {editor}
  {\bibfnamefont {Christoph}\ \bibnamefont {Koch}}\ and\ \bibinfo {editor}
  {\bibfnamefont {Idan}\ \bibnamefont {Segev}}}\ (\bibinfo  {publisher} {MIT
  Press},\ \bibinfo {year} {1998})\ pp.\ \bibinfo {pages} {251--291},\ \bibinfo
  {edition} {2nd}\ ed.\BibitemShut {Stop}%
\bibitem [{\citenamefont {Hindmarsh}\ and\ \citenamefont
  {Rose}(1984)}]{hindmarsh1984model}%
  \BibitemOpen
  \bibfield  {author} {\bibinfo {author} {\bibfnamefont {James~L.}\
  \bibnamefont {Hindmarsh}}\ and\ \bibinfo {author} {\bibfnamefont {R.M.}\
  \bibnamefont {Rose}},\ }\bibfield  {title} {\enquote {\bibinfo {title} {A
  model of neuronal bursting using three coupled first order differential
  equations},}\ }\href@noop {} {\bibfield  {journal} {\bibinfo  {journal}
  {{Proceedings of the Royal Society B: Biological Sciences}}\ }\textbf
  {\bibinfo {volume} {221}},\ \bibinfo {pages} {87--102} (\bibinfo {year}
  {1984})}\BibitemShut {NoStop}%
\bibitem [{\citenamefont {Izhikevich}(2007)}]{izhikevich_dynamical_2007}%
  \BibitemOpen
  \bibfield  {author} {\bibinfo {author} {\bibfnamefont {Eugene~M.}\
  \bibnamefont {Izhikevich}},\ }\href@noop {} {\emph {\bibinfo {title}
  {{Dynamical Systems in Neuroscience}}}}\ (\bibinfo  {publisher} {MIT Press},\
  \bibinfo {year} {2007})\BibitemShut {NoStop}%
\bibitem [{\citenamefont {Brette}(2013)}]{brette_sharpness_2013}%
  \BibitemOpen
  \bibfield  {author} {\bibinfo {author} {\bibfnamefont {Romain}\ \bibnamefont
  {Brette}},\ }\bibfield  {title} {\enquote {\bibinfo {title} {Sharpness of
  {Spike} {Initiation} in {Neurons} {Explained} by {Compartmentalization}},}\
  }\href {\doibase 10.1371/journal.pcbi.1003338} {\bibfield  {journal}
  {\bibinfo  {journal} {PLOS Comput Biol}\ }\textbf {\bibinfo {volume} {9}},\
  \bibinfo {pages} {e1003338} (\bibinfo {year} {2013})}\BibitemShut {NoStop}%
\bibitem [{\citenamefont {Naundorf}\ \emph {et~al.}(2006)\citenamefont
  {Naundorf}, \citenamefont {Wolf},\ and\ \citenamefont
  {Volgushev}}]{naundorf_unique_2006}%
  \BibitemOpen
  \bibfield  {author} {\bibinfo {author} {\bibfnamefont {Bj\"orn}\ \bibnamefont
  {Naundorf}}, \bibinfo {author} {\bibfnamefont {Fred}\ \bibnamefont {Wolf}}, \
  and\ \bibinfo {author} {\bibfnamefont {Maxim}\ \bibnamefont {Volgushev}},\
  }\bibfield  {title} {\enquote {\bibinfo {title} {Unique features of action
  potential initiation in cortical neurons},}\ }\href {\doibase
  10.1038/nature04610} {\bibfield  {journal} {\bibinfo  {journal} {Nature}\
  }\textbf {\bibinfo {volume} {440}},\ \bibinfo {pages} {1060--1063} (\bibinfo
  {year} {2006})}\BibitemShut {NoStop}%
\bibitem [{\citenamefont {Arhem}\ \emph {et~al.}(2006)\citenamefont {Arhem},
  \citenamefont {Klement},\ and\ \citenamefont
  {Blomberg}}]{arhem_channel_2006}%
  \BibitemOpen
  \bibfield  {author} {\bibinfo {author} {\bibfnamefont {Peter}\ \bibnamefont
  {Arhem}}, \bibinfo {author} {\bibfnamefont {G\:oran}\ \bibnamefont
  {Klement}}, \ and\ \bibinfo {author} {\bibfnamefont {Clas}\ \bibnamefont
  {Blomberg}},\ }\bibfield  {title} {\enquote {\bibinfo {title} {Channel
  {Density} {Regulation} of {Firing} {Patterns} in a {Cortical} {Neuron}
  {Model}},}\ }\href {\doibase 10.1529/biophysj.105.077032} {\bibfield
  {journal} {\bibinfo  {journal} {Biophysical Journal}\ }\textbf {\bibinfo
  {volume} {90}},\ \bibinfo {pages} {4392--4404} (\bibinfo {year}
  {2006})}\BibitemShut {NoStop}%
\bibitem [{\citenamefont {Stiefel}\ \emph {et~al.}(2008)\citenamefont
  {Stiefel}, \citenamefont {Gutkin},\ and\ \citenamefont
  {Sejnowski}}]{stiefel_effects_2008}%
  \BibitemOpen
  \bibfield  {author} {\bibinfo {author} {\bibfnamefont {Klaus~M.}\
  \bibnamefont {Stiefel}}, \bibinfo {author} {\bibfnamefont {Boris~S.}\
  \bibnamefont {Gutkin}}, \ and\ \bibinfo {author} {\bibfnamefont
  {Terrence~J.}\ \bibnamefont {Sejnowski}},\ }\bibfield  {title} {\enquote
  {\bibinfo {title} {The effects of cholinergic neuromodulation on neuronal
  phase-response curves of modeled cortical neurons},}\ }\href {\doibase
  10.1007/s10827-008-0111-9} {\bibfield  {journal} {\bibinfo  {journal}
  {Journal of {Computational} {Neuroscience}}\ }\textbf {\bibinfo {volume}
  {26}},\ \bibinfo {pages} {289--301} (\bibinfo {year} {2008})}\BibitemShut
  {NoStop}%
\bibitem [{\citenamefont {Prescott}\ \emph {et~al.}(2008)\citenamefont
  {Prescott}, \citenamefont {Ratt{\'e}}, \citenamefont {Koninck},\ and\
  \citenamefont {Sejnowski}}]{prescott_pyramidal_2008}%
  \BibitemOpen
  \bibfield  {author} {\bibinfo {author} {\bibfnamefont {Steven~A.}\
  \bibnamefont {Prescott}}, \bibinfo {author} {\bibfnamefont {St{\'e}phanie}\
  \bibnamefont {Ratt{\'e}}}, \bibinfo {author} {\bibfnamefont {Yves~De}\
  \bibnamefont {Koninck}}, \ and\ \bibinfo {author} {\bibfnamefont
  {Terrence~J.}\ \bibnamefont {Sejnowski}},\ }\bibfield  {title} {\enquote
  {\bibinfo {title} {Pyramidal {Neurons} {Switch} {From} {Integrators} {In}
  {Vitro} to {Resonators} {Under} {In} {Vivo}-{Like} {Conditions}},}\ }\href
  {\doibase 10.1152/jn.90634.2008} {\bibfield  {journal} {\bibinfo  {journal}
  {Journal of Neurophysiology}\ }\textbf {\bibinfo {volume} {100}},\ \bibinfo
  {pages} {3030--3042} (\bibinfo {year} {2008})}\BibitemShut {NoStop}%
\bibitem [{\citenamefont {Phoka}\ \emph {et~al.}(2010)\citenamefont {Phoka},
  \citenamefont {Cuntz}, \citenamefont {Roth},\ and\ \citenamefont
  {H{\:a}usser}}]{phoka_new_2010}%
  \BibitemOpen
  \bibfield  {author} {\bibinfo {author} {\bibfnamefont {Elena}\ \bibnamefont
  {Phoka}}, \bibinfo {author} {\bibfnamefont {Hermann}\ \bibnamefont {Cuntz}},
  \bibinfo {author} {\bibfnamefont {Arnd}\ \bibnamefont {Roth}}, \ and\
  \bibinfo {author} {\bibfnamefont {Michael}\ \bibnamefont {H{\:a}usser}},\
  }\bibfield  {title} {\enquote {\bibinfo {title} {A {New} {Approach} for
  {Determining} {Phase} {Response} {Curves} {Reveals} that {Purkinje} {Cells}
  {Can} {Act} as {Perfect} {Integrators}},}\ }\href {\doibase
  10.1371/journal.pcbi.1000768} {\bibfield  {journal} {\bibinfo  {journal}
  {PLoS Comput Biol}\ }\textbf {\bibinfo {volume} {6}},\ \bibinfo {pages}
  {e1000768} (\bibinfo {year} {2010})}\BibitemShut {NoStop}%
\bibitem [{\citenamefont {Richardson}\ \emph {et~al.}(2003)\citenamefont
  {Richardson}, \citenamefont {Brunel},\ and\ \citenamefont
  {Hakim}}]{richardson_subthreshold_2003}%
  \BibitemOpen
  \bibfield  {author} {\bibinfo {author} {\bibfnamefont {Magnus J.~E.}\
  \bibnamefont {Richardson}}, \bibinfo {author} {\bibfnamefont {Nicolas}\
  \bibnamefont {Brunel}}, \ and\ \bibinfo {author} {\bibfnamefont {Vincent}\
  \bibnamefont {Hakim}},\ }\bibfield  {title} {\enquote {\bibinfo {title} {From
  {Subthreshold} to {Firing}-{Rate} {Resonance}},}\ }\href {\doibase
  10.1152/jn.00955.2002} {\bibfield  {journal} {\bibinfo  {journal} {Journal of
  Neurophysiology}\ }\textbf {\bibinfo {volume} {89}},\ \bibinfo {pages}
  {2538--2554} (\bibinfo {year} {2003})}\BibitemShut {NoStop}%
\bibitem [{\citenamefont {Schreiber}\ \emph {et~al.}(2009)\citenamefont
  {Schreiber}, \citenamefont {Samengo},\ and\ \citenamefont
  {Herz}}]{schreiber_two_2009}%
  \BibitemOpen
  \bibfield  {author} {\bibinfo {author} {\bibfnamefont {Susanne}\ \bibnamefont
  {Schreiber}}, \bibinfo {author} {\bibfnamefont {In{\'e}s}\ \bibnamefont
  {Samengo}}, \ and\ \bibinfo {author} {\bibfnamefont {Andreas V.~M.}\
  \bibnamefont {Herz}},\ }\bibfield  {title} {\enquote {\bibinfo {title} {Two
  {Distinct} {Mechanisms} {Shape} the {Reliability} of {Neural} {Responses}},}\
  }\href {\doibase 10.1152/jn.90711.2008} {\bibfield  {journal} {\bibinfo
  {journal} {Journal of Neurophysiology}\ }\textbf {\bibinfo {volume} {101}},\
  \bibinfo {pages} {2239--2251} (\bibinfo {year} {2009})}\BibitemShut {NoStop}%
\bibitem [{\citenamefont {Schreiber}\ \emph {et~al.}(2004)\citenamefont
  {Schreiber}, \citenamefont {Erchova}, \citenamefont {Heinemann},\ and\
  \citenamefont {Herz}}]{schreiber_subthreshold_2004}%
  \BibitemOpen
  \bibfield  {author} {\bibinfo {author} {\bibfnamefont {Susanne}\ \bibnamefont
  {Schreiber}}, \bibinfo {author} {\bibfnamefont {Irina}\ \bibnamefont
  {Erchova}}, \bibinfo {author} {\bibfnamefont {Uwe}\ \bibnamefont
  {Heinemann}}, \ and\ \bibinfo {author} {\bibfnamefont {Andreas V.~M.}\
  \bibnamefont {Herz}},\ }\bibfield  {title} {\enquote {\bibinfo {title}
  {Subthreshold {Resonance} {Explains} the {Frequency}-{Dependent}
  {Integration} of {Periodic} as {Well} as {Random} {Stimuli} in the
  {Entorhinal} {Cortex}},}\ }\href {\doibase 10.1152/jn.01116.2003} {\bibfield
  {journal} {\bibinfo  {journal} {Journal of Neurophysiology}\ }\textbf
  {\bibinfo {volume} {92}},\ \bibinfo {pages} {408--415} (\bibinfo {year}
  {2004})}\BibitemShut {NoStop}%
\bibitem [{\citenamefont {Izhikevich}(2000)}]{izhikevich_neural_2000}%
  \BibitemOpen
  \bibfield  {author} {\bibinfo {author} {\bibfnamefont {E.}~\bibnamefont
  {Izhikevich}},\ }\bibfield  {title} {\enquote {\bibinfo {title} {Neural
  excitability, spiking and bursting},}\ }\href@noop {} {\bibfield  {journal}
  {\bibinfo  {journal} {International Journal of Bifurcation and Chaos}\
  }\textbf {\bibinfo {volume} {10}},\ \bibinfo {pages} {1171--1266} (\bibinfo
  {year} {2000})}\BibitemShut {NoStop}%
\bibitem [{\citenamefont {Schecter}(1987)}]{schecter_saddle-node_1987}%
  \BibitemOpen
  \bibfield  {author} {\bibinfo {author} {\bibfnamefont {Stephen}\ \bibnamefont
  {Schecter}},\ }\bibfield  {title} {\enquote {\bibinfo {title} {The
  {Saddle}-{Node} {Separatrix}-{Loop} {Bifurcation}},}\ }\href {\doibase
  10.1137/0518083} {\bibfield  {journal} {\bibinfo  {journal} {SIAM Journal on
  Mathematical Analysis}\ }\textbf {\bibinfo {volume} {18}},\ \bibinfo {pages}
  {1142--1156} (\bibinfo {year} {1987})}\BibitemShut {NoStop}%
\bibitem [{Note1()}]{Note1}%
  \BibitemOpen
  \bibinfo {note} {The SNL bifurcation \cite {schecter_saddle-node_1987} is
  also known as \protect \emph {saddle-node homoclinic orbit}, \protect \emph
  {saddle-node noncentral homoclinic}, \protect \emph {saddle-node
  separatrix-loop} bifurcation \cite {izhikevich_dynamical_2007} or \protect
  \emph {orbit flip} bifurcation \cite {homburg_homoclinic_2010}.}\BibitemShut
  {Stop}%
\bibitem [{\citenamefont {Jiruska}\ \emph {et~al.}(2013)\citenamefont
  {Jiruska}, \citenamefont {de~Curtis}, \citenamefont {Jefferys}, \citenamefont
  {Schevon}, \citenamefont {Schiff},\ and\ \citenamefont
  {Schindler}}]{jiruska_synchronization_2013}%
  \BibitemOpen
  \bibfield  {author} {\bibinfo {author} {\bibfnamefont {Premysl}\ \bibnamefont
  {Jiruska}}, \bibinfo {author} {\bibfnamefont {Marco}\ \bibnamefont
  {de~Curtis}}, \bibinfo {author} {\bibfnamefont {John G.~R.}\ \bibnamefont
  {Jefferys}}, \bibinfo {author} {\bibfnamefont {Catherine~A.}\ \bibnamefont
  {Schevon}}, \bibinfo {author} {\bibfnamefont {Steven~J.}\ \bibnamefont
  {Schiff}}, \ and\ \bibinfo {author} {\bibfnamefont {Kaspar}\ \bibnamefont
  {Schindler}},\ }\bibfield  {title} {\enquote {\bibinfo {title}
  {Synchronization and desynchronization in epilepsy: controversies and
  hypotheses},}\ }\href {\doibase 10.1113/jphysiol.2012.239590} {\bibfield
  {journal} {\bibinfo  {journal} {The Journal of Physiology}\ }\textbf
  {\bibinfo {volume} {591}},\ \bibinfo {pages} {787--797} (\bibinfo {year}
  {2013})}\BibitemShut {NoStop}%
\bibitem [{\citenamefont {Schiff}(2010)}]{schiff_towards_2010}%
  \BibitemOpen
  \bibfield  {author} {\bibinfo {author} {\bibfnamefont {Steven~J.}\
  \bibnamefont {Schiff}},\ }\bibfield  {title} {\enquote {\bibinfo {title}
  {Towards model-based control of {Parkinson}'s disease},}\ }\href {\doibase
  10.1098/rsta.2010.0050} {\bibfield  {journal} {\bibinfo  {journal}
  {{Philosophical Transactions of the Royal Society A: Mathematical, Physical
  and Engineering Sciences}}\ }\textbf {\bibinfo {volume} {368}},\ \bibinfo
  {pages} {2269--2308} (\bibinfo {year} {2010})}\BibitemShut {NoStop}%
\bibitem [{\citenamefont {Kirst}\ \emph {et~al.}(2015)\citenamefont {Kirst},
  \citenamefont {Ammer}, \citenamefont {Felmy}, \citenamefont {Herz},\ and\
  \citenamefont {Stemmler}}]{kirst_fundamental_2015}%
  \BibitemOpen
  \bibfield  {author} {\bibinfo {author} {\bibfnamefont {Christoph}\
  \bibnamefont {Kirst}}, \bibinfo {author} {\bibfnamefont {Julian}\
  \bibnamefont {Ammer}}, \bibinfo {author} {\bibfnamefont {Felix}\ \bibnamefont
  {Felmy}}, \bibinfo {author} {\bibfnamefont {Andreas}\ \bibnamefont {Herz}}, \
  and\ \bibinfo {author} {\bibfnamefont {Martin}\ \bibnamefont {Stemmler}},\
  }\bibfield  {title} {\enquote {\bibinfo {title} {Fundamental structure and
  modulation of neuronal excitability: Synaptic control of coding, resonance,
  and network synchronization},}\ }\href {\doibase 10.1101/022475} {\bibfield
  {journal} {\bibinfo  {journal} {bioRxiv}\ } (\bibinfo {year} {2015}),\
  10.1101/022475}\BibitemShut {NoStop}%
\bibitem [{\citenamefont {Pereira}\ \emph {et~al.}(2015)\citenamefont
  {Pereira}, \citenamefont {Coullet},\ and\ \citenamefont
  {Tirapegui}}]{pereira_bogdanovtakens_2015}%
  \BibitemOpen
  \bibfield  {author} {\bibinfo {author} {\bibfnamefont {Ulises}\ \bibnamefont
  {Pereira}}, \bibinfo {author} {\bibfnamefont {Pierre}\ \bibnamefont
  {Coullet}}, \ and\ \bibinfo {author} {\bibfnamefont {Enrique}\ \bibnamefont
  {Tirapegui}},\ }\bibfield  {title} {\enquote {\bibinfo {title} {The
  {{Bogdanov}-{Takens}} {Normal} {Form}: {A} {Minimal} {Model} for {Single}
  {Neuron} {Dynamics}},}\ }\href {\doibase 10.3390/e17127850} {\bibfield
  {journal} {\bibinfo  {journal} {Entropy}\ }\textbf {\bibinfo {volume} {17}},\
  \bibinfo {pages} {7859--7874} (\bibinfo {year} {2015})}\BibitemShut {NoStop}%
\bibitem [{\citenamefont {Ermentrout}\ \emph {et~al.}(2007)\citenamefont
  {Ermentrout}, \citenamefont {Galán},\ and\ \citenamefont
  {Urban}}]{ermentrout_relating_2007}%
  \BibitemOpen
  \bibfield  {author} {\bibinfo {author} {\bibfnamefont {G.~Bard}\ \bibnamefont
  {Ermentrout}}, \bibinfo {author} {\bibfnamefont {Roberto~F.}\ \bibnamefont
  {Galán}}, \ and\ \bibinfo {author} {\bibfnamefont {Nathaniel~N.}\
  \bibnamefont {Urban}},\ }\bibfield  {title} {\enquote {\bibinfo {title}
  {Relating {Neural} {Dynamics} to {Neural} {Coding}},}\ }\href {\doibase
  10.1103/PhysRevLett.99.248103} {\bibfield  {journal} {\bibinfo  {journal}
  {Physical Review Letters}\ }\textbf {\bibinfo {volume} {99}} (\bibinfo {year}
  {2007}),\ 10.1103/PhysRevLett.99.248103}\BibitemShut {NoStop}%
\bibitem [{\citenamefont {Rieke}\ \emph {et~al.}(1999)\citenamefont {Rieke},
  \citenamefont {{Warland, David}}, \citenamefont {{Bialek, William}},\ and\
  \citenamefont {{de Ruyter van Steveninck, Rob}}}]{rieke_spikes:_1999}%
  \BibitemOpen
  \bibfield  {author} {\bibinfo {author} {\bibfnamefont {Fred}\ \bibnamefont
  {Rieke}}, \bibinfo {author} {\bibnamefont {{Warland, David}}}, \bibinfo
  {author} {\bibnamefont {{Bialek, William}}}, \ and\ \bibinfo {author}
  {\bibnamefont {{de Ruyter van Steveninck, Rob}}},\ }\href@noop {} {\emph
  {\bibinfo {title} {Spikes: {Exploring} the {Neural} {Code}}}}\ (\bibinfo
  {publisher} {MIT Press},\ \bibinfo {year} {1999})\BibitemShut {NoStop}%
\bibitem [{\citenamefont {Lazar}(2010)}]{lazar_population_2010}%
  \BibitemOpen
  \bibfield  {author} {\bibinfo {author} {\bibfnamefont {Aurel~A.}\
  \bibnamefont {Lazar}},\ }\bibfield  {title} {\enquote {\bibinfo {title}
  {Population {Encoding} {With} {Hodgkin}–{Huxley} {Neurons}},}\ }\href
  {\doibase 10.1109/TIT.2009.2037040} {\bibfield  {journal} {\bibinfo
  {journal} {IEEE Transactions on Information Theory}\ }\textbf {\bibinfo
  {volume} {56}},\ \bibinfo {pages} {821--837} (\bibinfo {year}
  {2010})}\BibitemShut {NoStop}%
\bibitem [{\citenamefont {Bordyugov}\ \emph {et~al.}(2015)\citenamefont
  {Bordyugov}, \citenamefont {Abraham}, \citenamefont {Granada}, \citenamefont
  {Rose}, \citenamefont {Imkeller}, \citenamefont {Kramer},\ and\ \citenamefont
  {Herzel}}]{bordyugov_tuning_2015}%
  \BibitemOpen
  \bibfield  {author} {\bibinfo {author} {\bibfnamefont {Grigory}\ \bibnamefont
  {Bordyugov}}, \bibinfo {author} {\bibfnamefont {Ute}\ \bibnamefont
  {Abraham}}, \bibinfo {author} {\bibfnamefont {Adrian}\ \bibnamefont
  {Granada}}, \bibinfo {author} {\bibfnamefont {Pia}\ \bibnamefont {Rose}},
  \bibinfo {author} {\bibfnamefont {Katharina}\ \bibnamefont {Imkeller}},
  \bibinfo {author} {\bibfnamefont {Achim}\ \bibnamefont {Kramer}}, \ and\
  \bibinfo {author} {\bibfnamefont {Hanspeter}\ \bibnamefont {Herzel}},\
  }\bibfield  {title} {\enquote {\bibinfo {title} {Tuning the phase of
  circadian entrainment},}\ }\href {\doibase 10.1098/rsif.2015.0282} {\bibfield
   {journal} {\bibinfo  {journal} {Journal of The Royal Society Interface}\
  }\textbf {\bibinfo {volume} {12}},\ \bibinfo {pages} {20150282} (\bibinfo
  {year} {2015})}\BibitemShut {NoStop}%
\bibitem [{\citenamefont {Hilditch}(2001)}]{hilditch_introduction_2001}%
  \BibitemOpen
  \bibfield  {author} {\bibinfo {author} {\bibfnamefont {Ron~W.}\ \bibnamefont
  {Hilditch}},\ }\href@noop {} {\emph {\bibinfo {title} {An {Introduction} to
  {Close} {Binary} {Stars}}}}\ (\bibinfo  {publisher} {Cambridge University
  Press},\ \bibinfo {year} {2001})\BibitemShut {NoStop}%
\bibitem [{\citenamefont {Lazar}\ and\ \citenamefont
  {Slutskiy}(2014)}]{lazar_functional_2014}%
  \BibitemOpen
  \bibfield  {author} {\bibinfo {author} {\bibfnamefont {Aurel~A.}\
  \bibnamefont {Lazar}}\ and\ \bibinfo {author} {\bibfnamefont {Yevgeniy~B.}\
  \bibnamefont {Slutskiy}},\ }\bibfield  {title} {\enquote {\bibinfo {title}
  {Functional {Identification} of {Spike}-{Processing} {Neural} {Circuits}},}\
  }\href {\doibase 10.1162/NECO_a_00543} {\bibfield  {journal} {\bibinfo
  {journal} {Neural Computation}\ }\textbf {\bibinfo {volume} {26}},\ \bibinfo
  {pages} {264--305} (\bibinfo {year} {2014})}\BibitemShut {NoStop}%
\bibitem [{\citenamefont {Kuznetsov}(2013)}]{kuznetsov_elements_2013}%
  \BibitemOpen
  \bibfield  {author} {\bibinfo {author} {\bibfnamefont {Yuri~A.}\ \bibnamefont
  {Kuznetsov}},\ }\href@noop {} {\emph {\bibinfo {title} {Elements of {Applied}
  {Bifurcation} {Theory}}}}\ (\bibinfo  {publisher} {Springer Science \&
  Business Media},\ \bibinfo {year} {2013})\BibitemShut {NoStop}%
\bibitem [{\citenamefont {Homburg}\ and\ \citenamefont
  {Sandstede}(2010)}]{homburg_homoclinic_2010}%
  \BibitemOpen
  \bibfield  {author} {\bibinfo {author} {\bibfnamefont {Ale~Jan}\ \bibnamefont
  {Homburg}}\ and\ \bibinfo {author} {\bibfnamefont {Bj{\"o}rn}\ \bibnamefont
  {Sandstede}},\ }\bibfield  {title} {\enquote {\bibinfo {title} {Homoclinic
  and heteroclinic bifurcations in vector fields},}\ }in\ \href@noop {} {\emph
  {\bibinfo {booktitle} {Handbook of Dynamical Systems {III}}}},\ \bibinfo
  {editor} {edited by\ \bibinfo {editor} {\bibfnamefont {Henk~W.}\ \bibnamefont
  {Broer}}, \bibinfo {editor} {\bibfnamefont {Floris}\ \bibnamefont {Takens}},
  \ and\ \bibinfo {editor} {\bibfnamefont {Boris}\ \bibnamefont
  {Hasselblatt}}}\ (\bibinfo  {publisher} {Elsevier},\ \bibinfo {year}
  {2010})\BibitemShut {NoStop}%
\bibitem [{\citenamefont {Brown}\ \emph {et~al.}(2004)\citenamefont {Brown},
  \citenamefont {Moehlis},\ and\ \citenamefont {Holmes}}]{brown_phase_2004}%
  \BibitemOpen
  \bibfield  {author} {\bibinfo {author} {\bibfnamefont {Eric}\ \bibnamefont
  {Brown}}, \bibinfo {author} {\bibfnamefont {Jeff}\ \bibnamefont {Moehlis}}, \
  and\ \bibinfo {author} {\bibfnamefont {Philip}\ \bibnamefont {Holmes}},\
  }\bibfield  {title} {\enquote {\bibinfo {title} {On the {Phase} {Reduction}
  and {Response} {Dynamics} of {Neural} {Oscillator} {Populations}},}\ }\href
  {\doibase 10.1162/089976604322860668} {\bibfield  {journal} {\bibinfo
  {journal} {Neural Computation}\ }\textbf {\bibinfo {volume} {16}},\ \bibinfo
  {pages} {673--715} (\bibinfo {year} {2004})}\BibitemShut {NoStop}%
\bibitem [{\citenamefont {Shaw}\ \emph {et~al.}(2012)\citenamefont {Shaw},
  \citenamefont {Park}, \citenamefont {Chiel},\ and\ \citenamefont
  {Thomas}}]{shaw_phase_2012}%
  \BibitemOpen
  \bibfield  {author} {\bibinfo {author} {\bibfnamefont {Kendrick~M.}\
  \bibnamefont {Shaw}}, \bibinfo {author} {\bibfnamefont {Young-Min}\
  \bibnamefont {Park}}, \bibinfo {author} {\bibfnamefont {Hillel~J.}\
  \bibnamefont {Chiel}}, \ and\ \bibinfo {author} {\bibfnamefont {Peter~J.}\
  \bibnamefont {Thomas}},\ }\bibfield  {title} {\enquote {\bibinfo {title}
  {Phase {Resetting} in an {Asymptotically} {Phaseless} {System}: {On} the
  {Phase} {Response} of {Limit} {Cycles} {Verging} on a {Heteroclinic}
  {Orbit}},}\ }\href {\doibase 10.1137/110828976} {\bibfield  {journal}
  {\bibinfo  {journal} {SIAM Journal on Applied Dynamical Systems}\ }\textbf
  {\bibinfo {volume} {11}},\ \bibinfo {pages} {350--391} (\bibinfo {year}
  {2012})}\BibitemShut {NoStop}%
\bibitem [{\citenamefont {Machens}\ \emph {et~al.}(2001)\citenamefont
  {Machens}, \citenamefont {Stemmler}, \citenamefont {Prinz}, \citenamefont
  {Krahe}, \citenamefont {Ronacher},\ and\ \citenamefont
  {Herz}}]{machens_representation_2001}%
  \BibitemOpen
  \bibfield  {author} {\bibinfo {author} {\bibfnamefont {Christian~K.}\
  \bibnamefont {Machens}}, \bibinfo {author} {\bibfnamefont {Martin~B.}\
  \bibnamefont {Stemmler}}, \bibinfo {author} {\bibfnamefont {Petra}\
  \bibnamefont {Prinz}}, \bibinfo {author} {\bibfnamefont {R{\"u}diger}\
  \bibnamefont {Krahe}}, \bibinfo {author} {\bibfnamefont {Bernhard}\
  \bibnamefont {Ronacher}}, \ and\ \bibinfo {author} {\bibfnamefont {Andreas
  V.~M.}\ \bibnamefont {Herz}},\ }\bibfield  {title} {\enquote {\bibinfo
  {title} {Representation of {Acoustic} {Communication} {Signals} by {Insect}
  {Auditory} {Receptor} {Neurons}},}\ }\href
  {http://www.jneurosci.org/content/21/9/3215} {\bibfield  {journal} {\bibinfo
  {journal} {The Journal of Neuroscience}\ }\textbf {\bibinfo {volume} {21}},\
  \bibinfo {pages} {3215--3227} (\bibinfo {year} {2001})}\BibitemShut {NoStop}%
\bibitem [{\citenamefont {Fox}\ \emph {et~al.}(2010)\citenamefont {Fox},
  \citenamefont {Fairhall},\ and\ \citenamefont {Daniel}}]{fox_encoding_2010}%
  \BibitemOpen
  \bibfield  {author} {\bibinfo {author} {\bibfnamefont {Jessica~L.}\
  \bibnamefont {Fox}}, \bibinfo {author} {\bibfnamefont {Adrienne~L.}\
  \bibnamefont {Fairhall}}, \ and\ \bibinfo {author} {\bibfnamefont
  {Thomas~L.}\ \bibnamefont {Daniel}},\ }\bibfield  {title} {\enquote {\bibinfo
  {title} {Encoding properties of haltere neurons enable motion feature
  detection in a biological gyroscope},}\ }\href {\doibase
  10.1073/pnas.0912548107} {\bibfield  {journal} {\bibinfo  {journal}
  {Proceedings of the National Academy of Sciences}\ }\textbf {\bibinfo
  {volume} {107}},\ \bibinfo {pages} {3840--3845} (\bibinfo {year}
  {2010})}\BibitemShut {NoStop}%
\bibitem [{\citenamefont {Neiman}\ and\ \citenamefont
  {Russell}(2011)}]{neiman_sensory_2011}%
  \BibitemOpen
  \bibfield  {author} {\bibinfo {author} {\bibfnamefont {Alexander~B.}\
  \bibnamefont {Neiman}}\ and\ \bibinfo {author} {\bibfnamefont {David~F.}\
  \bibnamefont {Russell}},\ }\bibfield  {title} {\enquote {\bibinfo {title}
  {Sensory coding in oscillatory electroreceptors of paddlefish},}\ }\href
  {\doibase 10.1063/1.3669494} {\bibfield  {journal} {\bibinfo  {journal}
  {Chaos: An Interdisciplinary Journal of Nonlinear Science}\ }\textbf
  {\bibinfo {volume} {21}},\ \bibinfo {pages} {047505} (\bibinfo {year}
  {2011})}\BibitemShut {NoStop}%
\bibitem [{\citenamefont {Cover}\ and\ \citenamefont
  {Thomas}(2012)}]{cover_elements_2012}%
  \BibitemOpen
  \bibfield  {author} {\bibinfo {author} {\bibfnamefont {Thomas~M.}\
  \bibnamefont {Cover}}\ and\ \bibinfo {author} {\bibfnamefont {Joy~A.}\
  \bibnamefont {Thomas}},\ }\href@noop {} {\emph {\bibinfo {title} {Elements of
  {Information} {Theory}}}}\ (\bibinfo  {publisher} {John Wiley \& Sons},\
  \bibinfo {year} {2012})\BibitemShut {NoStop}%
\bibitem [{\citenamefont {Chacron}\ \emph {et~al.}(2004)\citenamefont
  {Chacron}, \citenamefont {Lindner},\ and\ \citenamefont
  {Longtin}}]{chacron_noise_2004}%
  \BibitemOpen
  \bibfield  {author} {\bibinfo {author} {\bibfnamefont {Maurice~J.}\
  \bibnamefont {Chacron}}, \bibinfo {author} {\bibfnamefont {Benjamin}\
  \bibnamefont {Lindner}}, \ and\ \bibinfo {author} {\bibfnamefont {André}\
  \bibnamefont {Longtin}},\ }\bibfield  {title} {\enquote {\bibinfo {title}
  {Noise {Shaping} by {Interval} {Correlations} {Increases} {Information}
  {Transfer}},}\ }\href {\doibase 10.1103/PhysRevLett.92.080601} {\bibfield
  {journal} {\bibinfo  {journal} {Physical Review Letters}\ }\textbf {\bibinfo
  {volume} {92}},\ \bibinfo {pages} {080601} (\bibinfo {year}
  {2004})}\BibitemShut {NoStop}%
\bibitem [{\citenamefont {Mainen}\ and\ \citenamefont
  {Sejnowski}(1995)}]{mainen_reliability_1995}%
  \BibitemOpen
  \bibfield  {author} {\bibinfo {author} {\bibfnamefont {Zachary~F.}\
  \bibnamefont {Mainen}}\ and\ \bibinfo {author} {\bibfnamefont {Terrence~J.}\
  \bibnamefont {Sejnowski}},\ }\bibfield  {title} {\enquote {\bibinfo {title}
  {Reliability of spike timing in neocortical neurons},}\ }\href {\doibase
  10.1126/science.7770778} {\bibfield  {journal} {\bibinfo  {journal}
  {Science}\ }\textbf {\bibinfo {volume} {268}},\ \bibinfo {pages} {1503--1506}
  (\bibinfo {year} {1995})}\BibitemShut {NoStop}%
\bibitem [{\citenamefont {Pikovsky}\ \emph {et~al.}(2003)\citenamefont
  {Pikovsky}, \citenamefont {Rosenblum},\ and\ \citenamefont
  {Kurths}}]{pikovsky_synchronization:_2003}%
  \BibitemOpen
  \bibfield  {author} {\bibinfo {author} {\bibfnamefont {Arkady}\ \bibnamefont
  {Pikovsky}}, \bibinfo {author} {\bibfnamefont {Michael}\ \bibnamefont
  {Rosenblum}}, \ and\ \bibinfo {author} {\bibfnamefont {J{\"u}rgen}\
  \bibnamefont {Kurths}},\ }\href@noop {} {\emph {\bibinfo {title}
  {Synchronization: {A} {Universal} {Concept} in {Nonlinear} {Sciences}}}}\
  (\bibinfo  {publisher} {Cambridge University Press},\ \bibinfo {year}
  {2003})\BibitemShut {NoStop}%
\bibitem [{\citenamefont {Marella}\ and\ \citenamefont
  {Ermentrout}(2008)}]{marella_class-ii_2008}%
  \BibitemOpen
  \bibfield  {author} {\bibinfo {author} {\bibfnamefont {Sashi}\ \bibnamefont
  {Marella}}\ and\ \bibinfo {author} {\bibfnamefont {G.~Bard}\ \bibnamefont
  {Ermentrout}},\ }\bibfield  {title} {\enquote {\bibinfo {title} {Class-{II}
  neurons display a higher degree of stochastic synchronization than class-{I}
  neurons},}\ }\href {\doibase 10.1103/PhysRevE.77.041918} {\bibfield
  {journal} {\bibinfo  {journal} {Physical Review E}\ }\textbf {\bibinfo
  {volume} {77}},\ \bibinfo {pages} {041918} (\bibinfo {year}
  {2008})}\BibitemShut {NoStop}%
\bibitem [{Note2()}]{Note2}%
  \BibitemOpen
  \bibinfo {note} {Assuming the phase difference $\psi =\phi _1-\phi _2$ of two
  phase oscillaotors (Eq.~\protect \textup {\hbox {\mathsurround \z@ \protect
  \normalfont (\ignorespaces \ref {eq.noisy_phaseoscil}\unskip \@@italiccorr
  )}} with common input $I(t)$ and individual noise $\xi (t)$) evolves slower
  than $\phi _{1,2}$ itself, averaging yields a Fokker-Planck equation for the
  phase difference: $\protect \mathaccentV {dot}05Fp(\psi )=\protect \frac
  12\sigma ^2p''(\psi ) -\lambda (\psi p'(\psi )+p(\psi ))$. The linear
  coefficient can be calculated by the Novikov-Furutzu-Donsker formula $\lambda
  =<Z'(\phi _2)x(t)>=-\varepsilon ^2\DOTSI \intop \ilimits@ _0^1\protect \text
  d\phi \protect \tmspace +\thinmuskip {.1667em}(Z'(\phi ))^2$ \cite
  {teramae_robustness_2004,goldobin_antireliability_2006}. The steady density
  of the phase difference has variance $\lambda /\sigma ^2$, which is related
  to the inverse of a spike metric. All other eigenfunctions are Hermitian
  polynomials that decay with $(k\lambda )^{-1}$ \cite
  {gardiner_handbook_2004}. Both the decay to the stationary density and its
  variance are influence by the derivative of the PRC \cite
  {marella_class-ii_2008}.}\BibitemShut {Stop}%
\bibitem [{\citenamefont {Kuramoto}(1984)}]{kuramoto_chemical_1984}%
  \BibitemOpen
  \bibfield  {author} {\bibinfo {author} {\bibfnamefont {Yoshiki}\ \bibnamefont
  {Kuramoto}},\ }\href@noop {} {\emph {\bibinfo {title} {Chemical
  {Oscillations}, {Waves}, and {Turbulence}}}}\ (\bibinfo  {publisher}
  {Springer Science \& Business Media},\ \bibinfo {year} {1984})\BibitemShut
  {NoStop}%
\bibitem [{\citenamefont {Ermentrout}\ and\ \citenamefont
  {Terman}(2010)}]{ermentrout_mathematical_2010}%
  \BibitemOpen
  \bibfield  {author} {\bibinfo {author} {\bibfnamefont {G.~Bard}\ \bibnamefont
  {Ermentrout}}\ and\ \bibinfo {author} {\bibfnamefont {David~H.}\ \bibnamefont
  {Terman}},\ }\href@noop {} {\emph {\bibinfo {title} {Mathematical
  {Foundations} of {Neuroscience}}}}\ (\bibinfo  {publisher} {Springer Science
  \& Business Media},\ \bibinfo {year} {2010})\BibitemShut {NoStop}%
\bibitem [{\citenamefont {Daido}(1992)}]{daido_order_1992}%
  \BibitemOpen
  \bibfield  {author} {\bibinfo {author} {\bibfnamefont {Hiroaki}\ \bibnamefont
  {Daido}},\ }\bibfield  {title} {\enquote {\bibinfo {title} {Order function
  and macroscopic mutual entrainment in uniformly coupled limit-cycle
  oscillators},}\ }\href
  {http://ptp.oxfordjournals.org/content/88/6/1213.short} {\bibfield  {journal}
  {\bibinfo  {journal} {Progress of theoretical physics}\ }\textbf {\bibinfo
  {volume} {88}},\ \bibinfo {pages} {1213--1218} (\bibinfo {year}
  {1992})}\BibitemShut {NoStop}%
\bibitem [{\citenamefont {Daido}(1996)}]{daido_onset_1996}%
  \BibitemOpen
  \bibfield  {author} {\bibinfo {author} {\bibfnamefont {Hiroaki}\ \bibnamefont
  {Daido}},\ }\bibfield  {title} {\enquote {\bibinfo {title} {Onset of
  cooperative entrainment in limit-cycle oscillators with uniform all-to-all
  interactions: bifurcation of the order function},}\ }\href {\doibase
  10.1016/0167-2789(95)00260-X} {\bibfield  {journal} {\bibinfo  {journal}
  {Physica D: Nonlinear Phenomena}\ }\textbf {\bibinfo {volume} {91}},\
  \bibinfo {pages} {24--66} (\bibinfo {year} {1996})}\BibitemShut {NoStop}%
\bibitem [{\citenamefont {Hata}\ \emph {et~al.}(2011)\citenamefont {Hata},
  \citenamefont {Arai}, \citenamefont {Gal{\'a}n},\ and\ \citenamefont
  {Nakao}}]{hata_optimal_2011}%
  \BibitemOpen
  \bibfield  {author} {\bibinfo {author} {\bibfnamefont {Shigefumi}\
  \bibnamefont {Hata}}, \bibinfo {author} {\bibfnamefont {Kensuke}\
  \bibnamefont {Arai}}, \bibinfo {author} {\bibfnamefont {Roberto~F.}\
  \bibnamefont {Gal{\'a}n}}, \ and\ \bibinfo {author} {\bibfnamefont {Hiroya}\
  \bibnamefont {Nakao}},\ }\bibfield  {title} {\enquote {\bibinfo {title}
  {Optimal phase response curves for stochastic synchronization of limit-cycle
  oscillators by common {Poisson} noise},}\ }\href {\doibase
  10.1103/PhysRevE.84.016229} {\bibfield  {journal} {\bibinfo  {journal}
  {Physical Review E}\ }\textbf {\bibinfo {volume} {84}},\ \bibinfo {pages}
  {016229} (\bibinfo {year} {2011})}\BibitemShut {NoStop}%
\bibitem [{\citenamefont {Friedrich}\ \emph {et~al.}(2004)\citenamefont
  {Friedrich}, \citenamefont {Habermann},\ and\ \citenamefont
  {Laurent}}]{friedrich_multiplexing_2004}%
  \BibitemOpen
  \bibfield  {author} {\bibinfo {author} {\bibfnamefont {Rainer~W.}\
  \bibnamefont {Friedrich}}, \bibinfo {author} {\bibfnamefont {Christopher~J.}\
  \bibnamefont {Habermann}}, \ and\ \bibinfo {author} {\bibfnamefont {Gilles}\
  \bibnamefont {Laurent}},\ }\bibfield  {title} {\enquote {\bibinfo {title}
  {Multiplexing using synchrony in the zebrafish olfactory bulb},}\ }\href
  {\doibase 10.1038/nn1292} {\bibfield  {journal} {\bibinfo  {journal} {Nature
  Neuroscience}\ }\textbf {\bibinfo {volume} {7}},\ \bibinfo {pages} {862--871}
  (\bibinfo {year} {2004})}\BibitemShut {NoStop}%
\bibitem [{\citenamefont {Franci}\ \emph {et~al.}(2013)\citenamefont {Franci},
  \citenamefont {Drion}, \citenamefont {Seutin},\ and\ \citenamefont
  {Sepulchre}}]{franci_balance_2013}%
  \BibitemOpen
  \bibfield  {author} {\bibinfo {author} {\bibfnamefont {Alessio}\ \bibnamefont
  {Franci}}, \bibinfo {author} {\bibfnamefont {Guillaume}\ \bibnamefont
  {Drion}}, \bibinfo {author} {\bibfnamefont {Vincent}\ \bibnamefont {Seutin}},
  \ and\ \bibinfo {author} {\bibfnamefont {Rodolphe}\ \bibnamefont
  {Sepulchre}},\ }\bibfield  {title} {\enquote {\bibinfo {title} {A {Balance}
  {Equation} {Determines} a {Switch} in {Neuronal} {Excitability}},}\ }\href
  {\doibase 10.1371/journal.pcbi.1003040} {\bibfield  {journal} {\bibinfo
  {journal} {PLoS Computational Biology}\ }\textbf {\bibinfo {volume} {9}},\
  \bibinfo {pages} {e1003040} (\bibinfo {year} {2013})}\BibitemShut {NoStop}%
\bibitem [{\citenamefont {De~Maesschalck}\ and\ \citenamefont
  {Wechselberger}(2015)}]{de_maesschalck_neural_2015}%
  \BibitemOpen
  \bibfield  {author} {\bibinfo {author} {\bibfnamefont {Peter}\ \bibnamefont
  {De~Maesschalck}}\ and\ \bibinfo {author} {\bibfnamefont {Martin}\
  \bibnamefont {Wechselberger}},\ }\bibfield  {title} {\enquote {\bibinfo
  {title} {Neural {Excitability} and {Singular} {Bifurcations}},}\ }\href
  {\doibase 10.1186/s13408-015-0029-2} {\bibfield  {journal} {\bibinfo
  {journal} {Journal of Mathematical Neuroscience}\ }\textbf {\bibinfo {volume}
  {5}},\ \bibinfo {pages} {16} (\bibinfo {year} {2015})}\BibitemShut {NoStop}%
\bibitem [{\citenamefont {Ostojic}\ \emph {et~al.}(2015)\citenamefont
  {Ostojic}, \citenamefont {Szapiro}, \citenamefont {Schwartz}, \citenamefont
  {Barbour}, \citenamefont {Brunel},\ and\ \citenamefont
  {Hakim}}]{ostojic_neuronal_2015}%
  \BibitemOpen
  \bibfield  {author} {\bibinfo {author} {\bibfnamefont {Srdjan}\ \bibnamefont
  {Ostojic}}, \bibinfo {author} {\bibfnamefont {Germ{\'a}n}\ \bibnamefont
  {Szapiro}}, \bibinfo {author} {\bibfnamefont {Eric}\ \bibnamefont
  {Schwartz}}, \bibinfo {author} {\bibfnamefont {Boris}\ \bibnamefont
  {Barbour}}, \bibinfo {author} {\bibfnamefont {Nicolas}\ \bibnamefont
  {Brunel}}, \ and\ \bibinfo {author} {\bibfnamefont {Vincent}\ \bibnamefont
  {Hakim}},\ }\bibfield  {title} {\enquote {\bibinfo {title} {Neuronal
  {Morphology} {Generates} {High}-{Frequency} {Firing} {Resonance}},}\ }\href
  {\doibase 10.1523/JNEUROSCI.3924-14.2015} {\bibfield  {journal} {\bibinfo
  {journal} {The Journal of Neuroscience}\ }\textbf {\bibinfo {volume} {35}},\
  \bibinfo {pages} {7056--7068} (\bibinfo {year} {2015})}\BibitemShut {NoStop}%
\bibitem [{\citenamefont {Fourcaud}\ and\ \citenamefont
  {Brunel}(2002)}]{fourcaud_dynamics_2002}%
  \BibitemOpen
  \bibfield  {author} {\bibinfo {author} {\bibfnamefont {Nicolas}\ \bibnamefont
  {Fourcaud}}\ and\ \bibinfo {author} {\bibfnamefont {Nicolas}\ \bibnamefont
  {Brunel}},\ }\bibfield  {title} {\enquote {\bibinfo {title} {Dynamics of the
  firing probability of noisy integrate-and-fire neurons},}\ }\href {\doibase
  10.1162/089976602320264015} {\bibfield  {journal} {\bibinfo  {journal}
  {Neural Computation}\ }\textbf {\bibinfo {volume} {14}},\ \bibinfo {pages}
  {2057--2110} (\bibinfo {year} {2002})}\BibitemShut {NoStop}%
\bibitem [{\citenamefont {Stein}\ \emph {et~al.}(1972)\citenamefont {Stein},
  \citenamefont {French},\ and\ \citenamefont
  {Holden}}]{stein1972j:coherenceInfo}%
  \BibitemOpen
  \bibfield  {author} {\bibinfo {author} {\bibfnamefont {Richard~B.}\
  \bibnamefont {Stein}}, \bibinfo {author} {\bibfnamefont {Andrew~S.}\
  \bibnamefont {French}}, \ and\ \bibinfo {author} {\bibfnamefont {Andrew~V.}\
  \bibnamefont {Holden}},\ }\bibfield  {title} {\enquote {\bibinfo {title} {The
  frequency response, coherence, and information capacity of two neuronal
  models},}\ }\href
  {http://linkinghub.elsevier.com/retrieve/pii/S0006349572860879} {\bibfield
  {journal} {\bibinfo  {journal} {Biophysical Journal}\ }\textbf {\bibinfo
  {volume} {12}},\ \bibinfo {pages} {295--322} (\bibinfo {year}
  {1972})}\BibitemShut {NoStop}%
\bibitem [{\citenamefont {Chow}\ and\ \citenamefont
  {White}(1996)}]{chow_spontaneous_1996}%
  \BibitemOpen
  \bibfield  {author} {\bibinfo {author} {\bibfnamefont {Carson~C.}\
  \bibnamefont {Chow}}\ and\ \bibinfo {author} {\bibfnamefont {John~A.}\
  \bibnamefont {White}},\ }\bibfield  {title} {\enquote {\bibinfo {title}
  {Spontaneous action potentials due to channel fluctuations},}\ }\href
  {\doibase 10.1016/S0006-3495(96)79494-8} {\bibfield  {journal} {\bibinfo
  {journal} {Biophysical Journal}\ }\textbf {\bibinfo {volume} {71}},\ \bibinfo
  {pages} {3013--3021} (\bibinfo {year} {1996})}\BibitemShut {NoStop}%
\bibitem [{\citenamefont {Hodgkin}\ and\ \citenamefont
  {Huxley}(1952)}]{hodgkin_quantitative_1952}%
  \BibitemOpen
  \bibfield  {author} {\bibinfo {author} {\bibfnamefont {Alan~L.}\ \bibnamefont
  {Hodgkin}}\ and\ \bibinfo {author} {\bibfnamefont {Andrew~F.}\ \bibnamefont
  {Huxley}},\ }\bibfield  {title} {\enquote {\bibinfo {title} {A quantitative
  description of membrane current and its application to conduction and
  excitation in nerve},}\ }\href
  {http://www.ncbi.nlm.nih.gov/pmc/articles/PMC1392413/} {\bibfield  {journal}
  {\bibinfo  {journal} {The {Journal} of {Physiology}}\ }\textbf {\bibinfo
  {volume} {117}},\ \bibinfo {pages} {500--544} (\bibinfo {year}
  {1952})}\BibitemShut {NoStop}%
\bibitem [{Note3()}]{Note3}%
  \BibitemOpen
  \bibinfo {note} {The bifurcation structure of the HH model arises in our
  model when the two fold bifurcations collide in a cusp bifurcation. This can
  be achieved by parameters that affect the shape of the nullclines, but is not
  possible with input and capacitance as bifurcation parameters as used
  here.}\BibitemShut {Stop}%
\bibitem [{Note4()}]{Note4}%
  \BibitemOpen
  \bibinfo {note} {The normal form used to calculate PRCs for the generalized
  Hopf bifurcation in Ref.~ \cite {brown_phase_2004} assumes a circular
  symmetric fold of limit cycles bifurcation. This holds locally around the
  subcritical Hopf bifurcation, but is unrealistic for full-blown pulse-like
  spikes, for which a separation of time scales is required in the
  dynamics.}\BibitemShut {Stop}%
\bibitem [{\citenamefont {Tsubo}\ \emph {et~al.}(2007)\citenamefont {Tsubo},
  \citenamefont {Teramae},\ and\ \citenamefont
  {Fukai}}]{tsubo_synchronization_2007}%
  \BibitemOpen
  \bibfield  {author} {\bibinfo {author} {\bibfnamefont {Yasuhiro}\
  \bibnamefont {Tsubo}}, \bibinfo {author} {\bibfnamefont {Jun-nosuke}\
  \bibnamefont {Teramae}}, \ and\ \bibinfo {author} {\bibfnamefont {Tomoki}\
  \bibnamefont {Fukai}},\ }\bibfield  {title} {\enquote {\bibinfo {title}
  {Synchronization of {Excitatory} {Neurons} with {Strongly} {Heterogeneous}
  {Phase} {Responses}},}\ }\href {\doibase 10.1103/PhysRevLett.99.228101}
  {\bibfield  {journal} {\bibinfo  {journal} {Physical Review Letters}\
  }\textbf {\bibinfo {volume} {99}},\ \bibinfo {pages} {228101} (\bibinfo
  {year} {2007})}\BibitemShut {NoStop}%
\bibitem [{\citenamefont {Abouzeid}\ and\ \citenamefont
  {Ermentrout}(2011)}]{abouzeid_correlation_2011}%
  \BibitemOpen
  \bibfield  {author} {\bibinfo {author} {\bibfnamefont {Aushra}\ \bibnamefont
  {Abouzeid}}\ and\ \bibinfo {author} {\bibfnamefont {Bard}\ \bibnamefont
  {Ermentrout}},\ }\bibfield  {title} {\enquote {\bibinfo {title} {Correlation
  transfer in stochastically driven neural oscillators over long and short time
  scales},}\ }\href {\doibase 10.1103/PhysRevE.84.061914} {\bibfield  {journal}
  {\bibinfo  {journal} {Physical Review E}\ }\textbf {\bibinfo {volume} {84}},\
  \bibinfo {pages} {061914} (\bibinfo {year} {2011})}\BibitemShut {NoStop}%
\bibitem [{\citenamefont {Stiefel}\ \emph {et~al.}(2009)\citenamefont
  {Stiefel}, \citenamefont {Gutkin},\ and\ \citenamefont
  {Sejnowski}}]{stiefel_effects_2009}%
  \BibitemOpen
  \bibfield  {author} {\bibinfo {author} {\bibfnamefont {Klaus~M.}\
  \bibnamefont {Stiefel}}, \bibinfo {author} {\bibfnamefont {Boris~S.}\
  \bibnamefont {Gutkin}}, \ and\ \bibinfo {author} {\bibfnamefont
  {Terrence~J.}\ \bibnamefont {Sejnowski}},\ }\bibfield  {title} {\enquote
  {\bibinfo {title} {The effects of cholinergic neuromodulation on neuronal
  phase-response curves of modeled cortical neurons},}\ }\href {\doibase
  10.1007/s10827-008-0111-9} {\bibfield  {journal} {\bibinfo  {journal}
  {Journal of computational neuroscience}\ }\textbf {\bibinfo {volume} {26}},\
  \bibinfo {pages} {289--301} (\bibinfo {year} {2009})}\BibitemShut {NoStop}%
\bibitem [{\citenamefont {Barth}\ and\ \citenamefont
  {Poulet}(2012)}]{barth_experimental_2012}%
  \BibitemOpen
  \bibfield  {author} {\bibinfo {author} {\bibfnamefont {Alison~L.}\
  \bibnamefont {Barth}}\ and\ \bibinfo {author} {\bibfnamefont {James F.~A.}\
  \bibnamefont {Poulet}},\ }\bibfield  {title} {\enquote {\bibinfo {title}
  {Experimental evidence for sparse firing in the neocortex},}\ }\href
  {\doibase 10.1016/j.tins.2012.03.008} {\bibfield  {journal} {\bibinfo
  {journal} {Trends in Neurosciences}\ }\textbf {\bibinfo {volume} {35}},\
  \bibinfo {pages} {345--355} (\bibinfo {year} {2012})}\BibitemShut {NoStop}%
\bibitem [{\citenamefont {Hoppensteadt}\ and\ \citenamefont
  {Izhikevich}(1997)}]{hoppensteadt_weakly_1997}%
  \BibitemOpen
  \bibfield  {author} {\bibinfo {author} {\bibfnamefont {Frank~C.}\
  \bibnamefont {Hoppensteadt}}\ and\ \bibinfo {author} {\bibfnamefont
  {Eugene~M.}\ \bibnamefont {Izhikevich}},\ }\href@noop {} {\emph {\bibinfo
  {title} {Weakly {Connected} {Neural} {Networks}}}}\ (\bibinfo  {publisher}
  {Springer Science \& Business Media},\ \bibinfo {year} {1997})\ \bibinfo
  {note} {google-Books-ID: ubLfkGj21hkC}\BibitemShut {NoStop}%
\bibitem [{\citenamefont {Wells}\ \emph {et~al.}(2005)\citenamefont {Wells},
  \citenamefont {Kao}, \citenamefont {Mariappan}, \citenamefont {Albea},
  \citenamefont {Jansen}, \citenamefont {Konrad},\ and\ \citenamefont
  {Mahadevan-Jansen}}]{wells_optical_2005}%
  \BibitemOpen
  \bibfield  {author} {\bibinfo {author} {\bibfnamefont {Jonathon}\
  \bibnamefont {Wells}}, \bibinfo {author} {\bibfnamefont {Chris}\ \bibnamefont
  {Kao}}, \bibinfo {author} {\bibfnamefont {Karthik}\ \bibnamefont
  {Mariappan}}, \bibinfo {author} {\bibfnamefont {Jeffrey}\ \bibnamefont
  {Albea}}, \bibinfo {author} {\bibfnamefont {E.~Duco}\ \bibnamefont {Jansen}},
  \bibinfo {author} {\bibfnamefont {Peter}\ \bibnamefont {Konrad}}, \ and\
  \bibinfo {author} {\bibfnamefont {Anita}\ \bibnamefont {Mahadevan-Jansen}},\
  }\bibfield  {title} {\enquote {\bibinfo {title} {Optical stimulation of
  neural tissue in vivo},}\ }\href {\doibase 10.1364/OL.30.000504} {\bibfield
  {journal} {\bibinfo  {journal} {Optics Letters}\ }\textbf {\bibinfo {volume}
  {30}},\ \bibinfo {pages} {504--506} (\bibinfo {year} {2005})}\BibitemShut
  {NoStop}%
\bibitem [{\citenamefont {Shapiro}\ \emph {et~al.}(2012)\citenamefont
  {Shapiro}, \citenamefont {Homma}, \citenamefont {Villarreal}, \citenamefont
  {Richter},\ and\ \citenamefont {Bezanilla}}]{shapiro_infrared_2012}%
  \BibitemOpen
  \bibfield  {author} {\bibinfo {author} {\bibfnamefont {Mikhail~G.}\
  \bibnamefont {Shapiro}}, \bibinfo {author} {\bibfnamefont {Kazuaki}\
  \bibnamefont {Homma}}, \bibinfo {author} {\bibfnamefont {Sebastian}\
  \bibnamefont {Villarreal}}, \bibinfo {author} {\bibfnamefont {Claus-Peter}\
  \bibnamefont {Richter}}, \ and\ \bibinfo {author} {\bibfnamefont {Francisco}\
  \bibnamefont {Bezanilla}},\ }\bibfield  {title} {\enquote {\bibinfo {title}
  {Infrared light excites cells by changing their electrical capacitance},}\
  }\href {\doibase 10.1038/ncomms1742} {\bibfield  {journal} {\bibinfo
  {journal} {Nature Communications}\ }\textbf {\bibinfo {volume} {3}},\
  \bibinfo {pages} {736} (\bibinfo {year} {2012})}\BibitemShut {NoStop}%
\bibitem [{\citenamefont {Schultheiss}\ \emph {et~al.}(2011)\citenamefont
  {Schultheiss}, \citenamefont {Prinz},\ and\ \citenamefont
  {Butera}}]{schultheiss_phase_2011}%
  \BibitemOpen
  \bibfield  {author} {\bibinfo {author} {\bibfnamefont {Nathan~W.}\
  \bibnamefont {Schultheiss}}, \bibinfo {author} {\bibfnamefont {Astrid~A.}\
  \bibnamefont {Prinz}}, \ and\ \bibinfo {author} {\bibfnamefont {Robert~J.}\
  \bibnamefont {Butera}},\ }\href@noop {} {\emph {\bibinfo {title} {Phase
  {Response} {Curves} in {Neuroscience}: {Theory}, {Experiment}, and
  {Analysis}}}}\ (\bibinfo  {publisher} {Springer Science \& Business Media},\
  \bibinfo {year} {2011})\BibitemShut {NoStop}%
\bibitem [{\citenamefont {Wang}\ \emph {et~al.}(2013)\citenamefont {Wang},
  \citenamefont {Musharoff}, \citenamefont {Canavier},\ and\ \citenamefont
  {Gasparini}}]{wang_hippocampal_2013}%
  \BibitemOpen
  \bibfield  {author} {\bibinfo {author} {\bibfnamefont {Shuoguo}\ \bibnamefont
  {Wang}}, \bibinfo {author} {\bibfnamefont {Maximilian~M.}\ \bibnamefont
  {Musharoff}}, \bibinfo {author} {\bibfnamefont {Carmen~C.}\ \bibnamefont
  {Canavier}}, \ and\ \bibinfo {author} {\bibfnamefont {Sonia}\ \bibnamefont
  {Gasparini}},\ }\bibfield  {title} {\enquote {\bibinfo {title} {Hippocampal
  {CA}1 pyramidal neurons exhibit type 1 phase-response curves and type 1
  excitability},}\ }\href {\doibase 10.1152/jn.00721.2012} {\bibfield
  {journal} {\bibinfo  {journal} {Journal of Neurophysiology}\ }\textbf
  {\bibinfo {volume} {109}},\ \bibinfo {pages} {2757--2766} (\bibinfo {year}
  {2013})}\BibitemShut {NoStop}%
\bibitem [{\citenamefont {Gutkin}\ \emph {et~al.}(2005)\citenamefont {Gutkin},
  \citenamefont {Ermentrout},\ and\ \citenamefont
  {Reyes}}]{gutkin_phase-response_2005}%
  \BibitemOpen
  \bibfield  {author} {\bibinfo {author} {\bibfnamefont {Boris~S.}\
  \bibnamefont {Gutkin}}, \bibinfo {author} {\bibfnamefont {G.~Bard}\
  \bibnamefont {Ermentrout}}, \ and\ \bibinfo {author} {\bibfnamefont
  {Alex~D.}\ \bibnamefont {Reyes}},\ }\bibfield  {title} {\enquote {\bibinfo
  {title} {Phase-{Response} {Curves} {Give} the {Responses} of {Neurons} to
  {Transient} {Inputs}},}\ }\href {\doibase 10.1152/jn.00359.2004} {\bibfield
  {journal} {\bibinfo  {journal} {Journal of Neurophysiology}\ }\textbf
  {\bibinfo {volume} {94}},\ \bibinfo {pages} {1623--1635} (\bibinfo {year}
  {2005})}\BibitemShut {NoStop}%
\bibitem [{\citenamefont {Ermentrout}\ \emph {et~al.}(2011)\citenamefont
  {Ermentrout}, \citenamefont {Beverlin~{II}}, \citenamefont {Troyer},\ and\
  \citenamefont {Netoff}}]{ermentrout_variance_2011}%
  \BibitemOpen
  \bibfield  {author} {\bibinfo {author} {\bibfnamefont {G.~Bard}\ \bibnamefont
  {Ermentrout}}, \bibinfo {author} {\bibfnamefont {Bryce}\ \bibnamefont
  {Beverlin~{II}}}, \bibinfo {author} {\bibfnamefont {Todd}\ \bibnamefont
  {Troyer}}, \ and\ \bibinfo {author} {\bibfnamefont {Theoden~I.}\ \bibnamefont
  {Netoff}},\ }\bibfield  {title} {\enquote {\bibinfo {title} {The variance of
  phase-resetting curves},}\ }\href {\doibase 10.1007/s10827-010-0305-9}
  {\bibfield  {journal} {\bibinfo  {journal} {Journal of Computational
  Neuroscience}\ }\textbf {\bibinfo {volume} {31}},\ \bibinfo {pages}
  {185--197} (\bibinfo {year} {2011})}\BibitemShut {NoStop}%
\bibitem [{\citenamefont {Eguia}\ and\ \citenamefont
  {Mindlin}(1999)}]{eguia_semiconductor_1999}%
  \BibitemOpen
  \bibfield  {author} {\bibinfo {author} {\bibfnamefont {Manuel~C.}\
  \bibnamefont {Eguia}}\ and\ \bibinfo {author} {\bibfnamefont {Gabriel~B.}\
  \bibnamefont {Mindlin}},\ }\bibfield  {title} {\enquote {\bibinfo {title}
  {Semiconductor laser with optical feedback: {From} excitable to deterministic
  low-frequency fluctuations},}\ }\href {\doibase 10.1103/PhysRevE.60.1551}
  {\bibfield  {journal} {\bibinfo  {journal} {Physical Review E}\ }\textbf
  {\bibinfo {volume} {60}},\ \bibinfo {pages} {1551--1557} (\bibinfo {year}
  {1999})}\BibitemShut {NoStop}%
\bibitem [{\citenamefont {Martinez~Avila}\ \emph {et~al.}(2004)\citenamefont
  {Martinez~Avila}, \citenamefont {Cavalcante},\ and\ \citenamefont
  {Leite}}]{martinez_avila_experimental_2004}%
  \BibitemOpen
  \bibfield  {author} {\bibinfo {author} {\bibfnamefont {Jhon~F.}\ \bibnamefont
  {Martinez~Avila}}, \bibinfo {author} {\bibfnamefont {Hugo L. D. de~S.}\
  \bibnamefont {Cavalcante}}, \ and\ \bibinfo {author} {\bibfnamefont
  {J.~R.~Rios}\ \bibnamefont {Leite}},\ }\bibfield  {title} {\enquote {\bibinfo
  {title} {Experimental {Deterministic} {Coherence} {Resonance}},}\ }\href
  {\doibase 10.1103/PhysRevLett.93.144101} {\bibfield  {journal} {\bibinfo
  {journal} {Physical Review Letters}\ }\textbf {\bibinfo {volume} {93}},\
  \bibinfo {pages} {144101} (\bibinfo {year} {2004})}\BibitemShut {NoStop}%
\bibitem [{\citenamefont {Ek{\c{s}}io{\u{g}}lu}\ \emph
  {et~al.}(2013)\citenamefont {Ek{\c{s}}io{\u{g}}lu}, \citenamefont
  {M{\"u}stecapl{\i}o{\u{g}}lu},\ and\ \citenamefont
  {G{\"u}ven}}]{eksioglu_dissipative_2013}%
  \BibitemOpen
  \bibfield  {author} {\bibinfo {author} {\bibfnamefont {Yasa}\ \bibnamefont
  {Ek{\c{s}}io{\u{g}}lu}}, \bibinfo {author} {\bibfnamefont {{\"O}zg{\"u}r~E.}\
  \bibnamefont {M{\"u}stecapl{\i}o{\u{g}}lu}}, \ and\ \bibinfo {author}
  {\bibfnamefont {Kaan}\ \bibnamefont {G{\"u}ven}},\ }\bibfield  {title}
  {\enquote {\bibinfo {title} {Dissipative {Josephson} junction of an optical
  soliton and a surface plasmon},}\ }\href {\doibase
  10.1103/PhysRevA.87.023823} {\bibfield  {journal} {\bibinfo  {journal}
  {Physical Review A}\ }\textbf {\bibinfo {volume} {87}},\ \bibinfo {pages}
  {023823} (\bibinfo {year} {2013})}\BibitemShut {NoStop}%
\bibitem [{\citenamefont {Labouvie}\ \emph {et~al.}(2016)\citenamefont
  {Labouvie}, \citenamefont {Santra}, \citenamefont {Heun},\ and\ \citenamefont
  {Ott}}]{labouvie_bistability_2016}%
  \BibitemOpen
  \bibfield  {author} {\bibinfo {author} {\bibfnamefont {Ralf}\ \bibnamefont
  {Labouvie}}, \bibinfo {author} {\bibfnamefont {Bodhaditya}\ \bibnamefont
  {Santra}}, \bibinfo {author} {\bibfnamefont {Simon}\ \bibnamefont {Heun}}, \
  and\ \bibinfo {author} {\bibfnamefont {Herwig}\ \bibnamefont {Ott}},\
  }\bibfield  {title} {\enquote {\bibinfo {title} {Bistability in a
  {Driven}-{Dissipative} {Superfluid}},}\ }\href {\doibase
  10.1103/PhysRevLett.116.235302} {\bibfield  {journal} {\bibinfo  {journal}
  {Physical Review Letters}\ }\textbf {\bibinfo {volume} {116}},\ \bibinfo
  {pages} {235302} (\bibinfo {year} {2016})}\BibitemShut {NoStop}%
\bibitem [{\citenamefont {Shimizu}\ \emph {et~al.}(1995)\citenamefont
  {Shimizu}, \citenamefont {Morooka},\ and\ \citenamefont
  {Morisue}}]{shimizu_relaxation_1995}%
  \BibitemOpen
  \bibfield  {author} {\bibinfo {author} {\bibfnamefont {Nobuhiro}\
  \bibnamefont {Shimizu}}, \bibinfo {author} {\bibfnamefont {Toshimitsu}\
  \bibnamefont {Morooka}}, \ and\ \bibinfo {author} {\bibfnamefont {Mititada}\
  \bibnamefont {Morisue}},\ }\bibfield  {title} {\enquote {\bibinfo {title}
  {Relaxation {Oscillator} {Using} {Nb}/{AlO}\_x/{Al}/{Nb} {Josephson}
  {Junctions} {Measured} with {DC} {Superconducting} {Quantum} {Interference}
  {Devices}},}\ }\href {\doibase 10.1143/JJAP.34.5588} {\bibfield  {journal}
  {\bibinfo  {journal} {Japanese {Journal} of {Applied} {Physics}}\ }\textbf
  {\bibinfo {volume} {34}},\ \bibinfo {pages} {5588--5591} (\bibinfo {year}
  {1995})}\BibitemShut {NoStop}%
\bibitem [{\citenamefont {Romanelli}\ \emph {et~al.}(2016)\citenamefont
  {Romanelli}, \citenamefont {Thorette}, \citenamefont {Brunel}, \citenamefont
  {Erneux},\ and\ \citenamefont {Vallet}}]{romanelli_excitable-like_2016}%
  \BibitemOpen
  \bibfield  {author} {\bibinfo {author} {\bibfnamefont {Marco}\ \bibnamefont
  {Romanelli}}, \bibinfo {author} {\bibfnamefont {Aur{\'e}lien}\ \bibnamefont
  {Thorette}}, \bibinfo {author} {\bibfnamefont {Marc}\ \bibnamefont {Brunel}},
  \bibinfo {author} {\bibfnamefont {Thomas}\ \bibnamefont {Erneux}}, \ and\
  \bibinfo {author} {\bibfnamefont {Marcand}\ \bibnamefont {Vallet}},\
  }\bibfield  {title} {\enquote {\bibinfo {title} {Excitable-like chaotic
  pulses in the bounded-phase regime of an opto-rf oscillator},}\ }\href
  {\doibase 10.1103/PhysRevA.94.043820} {\bibfield  {journal} {\bibinfo
  {journal} {Physical Review A}\ }\textbf {\bibinfo {volume} {94}},\ \bibinfo
  {pages} {043820} (\bibinfo {year} {2016})}\BibitemShut {NoStop}%
\bibitem [{\citenamefont {Schneider}\ and\ \citenamefont
  {M{\"u}nster}(1991)}]{schneider_chemical_1991}%
  \BibitemOpen
  \bibfield  {author} {\bibinfo {author} {\bibfnamefont {Friedemann~W.}\
  \bibnamefont {Schneider}}\ and\ \bibinfo {author} {\bibfnamefont {Arno~F.}\
  \bibnamefont {M{\"u}nster}},\ }\bibfield  {title} {\enquote {\bibinfo {title}
  {Chemical oscillations, chaos, and fluctuations in flow reactors},}\ }\href
  {\doibase 10.1021/j100159a012} {\bibfield  {journal} {\bibinfo  {journal}
  {The Journal of Physical Chemistry}\ }\textbf {\bibinfo {volume} {95}},\
  \bibinfo {pages} {2130--2138} (\bibinfo {year} {1991})}\BibitemShut {NoStop}%
\bibitem [{\citenamefont {Balakotaiah}\ \emph {et~al.}(1999)\citenamefont
  {Balakotaiah}, \citenamefont {Dommeti},\ and\ \citenamefont
  {Gupta}}]{balakotaiah_bifurcation_1999}%
  \BibitemOpen
  \bibfield  {author} {\bibinfo {author} {\bibfnamefont {Vemuri}\ \bibnamefont
  {Balakotaiah}}, \bibinfo {author} {\bibfnamefont {Sandra M.~S.}\ \bibnamefont
  {Dommeti}}, \ and\ \bibinfo {author} {\bibfnamefont {Nikunj}\ \bibnamefont
  {Gupta}},\ }\bibfield  {title} {\enquote {\bibinfo {title} {Bifurcation
  analysis of chemical reactors and reacting flows},}\ }\href {\doibase
  10.1063/1.166377} {\bibfield  {journal} {\bibinfo  {journal} {Chaos: {An}
  {Interdisciplinary} {Journal} of {Nonlinear} {Science}}\ }\textbf {\bibinfo
  {volume} {9}},\ \bibinfo {pages} {13--35} (\bibinfo {year}
  {1999})}\BibitemShut {NoStop}%
\bibitem [{\citenamefont {Izhikevich}\ and\ \citenamefont
  {Hoppensteadt}(2003)}]{izhikevich_slowly_2003}%
  \BibitemOpen
  \bibfield  {author} {\bibinfo {author} {\bibfnamefont {Eugene~M.}\
  \bibnamefont {Izhikevich}}\ and\ \bibinfo {author} {\bibfnamefont {Frank~C.}\
  \bibnamefont {Hoppensteadt}},\ }\bibfield  {title} {\enquote {\bibinfo
  {title} {Slowly {Coupled} {Oscillators}: {Phase} {Dynamics} and
  {Synchronization}},}\ }\href {\doibase 10.1137/S0036139902400945} {\bibfield
  {journal} {\bibinfo  {journal} {SIAM Journal on Applied Mathematics}\
  }\textbf {\bibinfo {volume} {63}},\ \bibinfo {pages} {1935--1953} (\bibinfo
  {year} {2003})}\BibitemShut {NoStop}%
\bibitem [{\citenamefont {Chicone}(2006)}]{chicone_ordinary_2006}%
  \BibitemOpen
  \bibfield  {author} {\bibinfo {author} {\bibfnamefont {Carmen}\ \bibnamefont
  {Chicone}},\ }\href@noop {} {\emph {\bibinfo {title} {Ordinary {Differential}
  {Equations} with {Applications}}}}\ (\bibinfo  {publisher} {Springer Science
  \& Business Media},\ \bibinfo {year} {2006})\BibitemShut {NoStop}%
\bibitem [{\citenamefont {Teramae}\ and\ \citenamefont
  {Tanaka}(2004)}]{teramae_robustness_2004}%
  \BibitemOpen
  \bibfield  {author} {\bibinfo {author} {\bibfnamefont {Jun-nosuke}\
  \bibnamefont {Teramae}}\ and\ \bibinfo {author} {\bibfnamefont {Dan}\
  \bibnamefont {Tanaka}},\ }\bibfield  {title} {\enquote {\bibinfo {title}
  {Robustness of the {Noise}-{Induced} {Phase} {Synchronization} in a {General}
  {Class} of {Limit} {Cycle} {Oscillators}},}\ }\href {\doibase
  10.1103/PhysRevLett.93.204103} {\bibfield  {journal} {\bibinfo  {journal}
  {Physical Review Letters}\ }\textbf {\bibinfo {volume} {93}},\ \bibinfo
  {pages} {204103} (\bibinfo {year} {2004})}\BibitemShut {NoStop}%
\bibitem [{\citenamefont {Goldobin}\ and\ \citenamefont
  {Pikovsky}(2006)}]{goldobin_antireliability_2006}%
  \BibitemOpen
  \bibfield  {author} {\bibinfo {author} {\bibfnamefont {Denis~S.}\
  \bibnamefont {Goldobin}}\ and\ \bibinfo {author} {\bibfnamefont {Arkady}\
  \bibnamefont {Pikovsky}},\ }\bibfield  {title} {\enquote {\bibinfo {title}
  {Antireliability of noise-driven neurons},}\ }\href {\doibase
  10.1103/PhysRevE.73.061906} {\bibfield  {journal} {\bibinfo  {journal}
  {Physical Review E}\ }\textbf {\bibinfo {volume} {73}},\ \bibinfo {pages}
  {061906} (\bibinfo {year} {2006})}\BibitemShut {NoStop}%
\bibitem [{\citenamefont {Gardiner}(2004)}]{gardiner_handbook_2004}%
  \BibitemOpen
  \bibfield  {author} {\bibinfo {author} {\bibfnamefont {Crispin~W.}\
  \bibnamefont {Gardiner}},\ }\href@noop {} {\emph {\bibinfo {title} {Handbook
  of stochastic methods for physics, chemistry, and the natural sciences}}},\
  \bibinfo {edition} {3rd}\ ed.,\ Springer series in synergetics\ (\bibinfo
  {publisher} {Springer-Verlag},\ \bibinfo {address} {Berlin ; New York},\
  \bibinfo {year} {2004})\BibitemShut {NoStop}%
\end{thebibliography}%

\end{document}